\documentclass[amsmath,twocolumn,pra,nofootinbib,longbibliography]{revtex4-2} 
\usepackage{amsthm,amssymb,amsfonts,graphicx,verbatim, xcolor,bm} 
\usepackage{hyperref} 
\usepackage[utf8]{inputenc} 
\usepackage[T1]{fontenc} 
\usepackage{siunitx} 

\let\oldmarginpar\marginpar
\renewcommand\marginpar[1]{\-\oldmarginpar[\raggedleft\footnotesize #1]%
{\raggedright\footnotesize #1}}
\renewcommand{\eqref}[1]{(\ref{#1})}

\renewcommand{\vec}[1]{{\bf #1}}

\renewcommand{\Re}{{\rm Re}\,}

\newcommand{\ket}[1]{|#1\rangle}

\begin{document}

\title{
Spin-triplet superconductivity from interband effect in doped insulators
}
\author{Valentin Cr\'epel, Liang Fu}
\affiliation{Massachusetts Institute of Technology, 77 Massachusetts Avenue, Cambridge, MA, USA}

\begin{abstract}
Despite being of fundamental importance and potential interest for topological quantum computing, spin-triplet superconductors remain rare in solid state materials after decades of research. 
In this work, we present a general mechanism for spin-triplet superconductivity in multi-band systems, where a non-retarded pairing interaction between conduction electrons is produced by their electronic repulsion to a third-electron undergoing a virtual interband transition. 
Our theory is analytically controlled by an interband hybridization parameter, and explicitly demonstrated in doped band insulators with the example of an extended Hubbard model. 
In light of this theory, we propose that recently discovered dilute superconductors such as ZrNCl and WTe$_2$ are spin-triplet, and compare the expected consequences of our theory with experimental data. 
\end{abstract}

\maketitle

\section{Introduction}

Spin-triplet superconductors display a plethora of unconventional phenomena, including multi-component order parameter, fractional vortices~\cite{vakaryuk2009spin,salomaa1987quantized}, Majorana fermions~\cite{read2000paired} and topological boundary modes~\cite{schnyder2008classification,hsieh2012majorana}.
Further interest in spin-triplet superconductors is fueled by their prospect as a material platform for topological qubits~\cite{alicea2012new,beenakker2013search,sarma2015majorana,crepel2019variational}. 
However, triplet superconductors are rare to find. 
Inspired by superfluid helium-3~\cite{leggett1975theoretical}, the search for triplet pairing has traditionally been focused on nearly ferromagnetic metals, such as Sr$_2$RuO$_4$~\cite{mackenzie2003superconductivity}, UPt$_3$~\cite{sauls1994order} and UTe$_2$~\cite{ran2019nearly,jiao2020chiral}. 
In recent years, evidence of anisotropic spin-triplet pairing has also been discovered in superconducting 
doped topological insulators such as Cu$_x$Bi$_2$Se$_3$~\cite{matano2016spin,yonezawa2017thermodynamic,willa2018nanocalorimetric,cho2020z}, where strong spin-orbit coupling plays an important role~\cite{fu2010odd,hashimoto2013bulk,PhysRevB.90.100509,wan2014turning}. 
Despite significant effort and progress, the pairing symmetry and/or pairing mechanism of these candidate materials remain to be fully understood.

In this work, we introduce an electronic mechanism for spin-triplet superconductivity in doped insulators, where the pairing of doped electrons arises from interband electronic effects.  
By developing a controlled expansion in the interband hybridization, we show that an attractive interaction between two conduction electrons arises from virtual interband transition, as illustrated in Fig.1a.   
Since this mechanism involves two electrons forming the pair and a third one undergoing a virtual interband transition, we dub it ``three-particle mechanism'' for superconductivity~\cite{2020arXiv201208528C}. 

The idea of using interband electronic excitations, such as excitons, as a replacement of phonon to mediate superconductivity has a long history~\cite{little1964possibility,ginzburg1972problem,allender1973model,hirsch1986enhanced}. However, 
experimental evidence of excion-mediated superconductivity remains elusive. 
A major challenge is that most proposals rely on metal layers on a separate excitonic medium, which often result in weak coupling between conduction electrons and virtual excitons. Moreover, theoretical works on this subject have only considered s-wave pairing, which is usually disfavored in electron systems with strong repulsive interaction. 
Last but not the least, the large energy scale of intermediate excited states means that the induced interaction 
is {\it non-retarded}, in contrast to phonon-mediated pairing. 
Thus, new theoretical methods are needed to tackle the problem of superconductivity from repulsive interaction 
in multiband systems.   

The novelties of our work solve the above problems and challenges. We study multiband systems that naturally host both 
excitons and conduction electrons, interacting strongly  with each other by electrostatic forces. 
Using a two-band Hubbard model as an example, we show with exact solutions 
that virtual interband effects lead to spin-triplet pairing of two doped electrons in a band insulator. 
The spin-triplet electron pair naturally avoids the large Coulomb repulsion at short distance.  

Since our hybridization expansion method is non-perturbative in the interaction strength,  
we obtain an asymptotically exact theory of strong-coupling superconductivity at low doping, 
without distractions from other competing states and without requiring any bosonic glue. 
Our theory provides a mechanism whereby superconductivity arises upon {\it infinitesimal} doping of a band insulator.  
It predicts a direct transition from a band insulator to a superconductor without single-particle gap closing, 
and a BEC/BCS crossover as a function of doping concentration. 



Our theory sheds light on unconventional superconductivity behaviors in doped band insulators. 
We shall focus on two materials: electron-doped ZrNCl and WTe$_2$, both of which superconduct at very low doping.  
Remarkably, BEC-BCS crossover has recently been observed in two-dimensional ZrNCl~\cite{nakagawa2020gate}. 
In monolayer WTe$_2$, a direct insulator-superconductor transition was found under electrostatic gating~\cite{fatemi2018electrically,sajadi2018gate}. 
We will compare our theoretical predictions of spin-triplet superconductivity with the experimental data on these dilute superconductors. 
In particular, we highlight the observed increase of $T_c$ in WTe$_2$ under a small in-plane magnetic field as 
a strong evidence for spin-triplet superconductivity.

\section{Model}

To illustrate the interband electronic mechanism for superconductivity, we consider a two-band Hubbard model on the honeycomb lattice with staggered potential on $A$/$B$ sites and extended interactions. The Hamiltonian takes the form   
\begin{align} \label{eq_original_model}
& \mathcal{H} = -t_0 \sum_{\langle r, r'\rangle, \sigma} ( c_{r,\sigma}^\dagger c_{r', \sigma} + hc ) + \frac{\Delta_0}{2} \left[ \sum_{r\in B} n_r  -\sum_{r\in A} n_r  \right] \nonumber   \\
&+ U_A \sum_{r\in A} n_{r\uparrow} n_{r\downarrow} + U_B \sum_{r\in B} n_{r\uparrow} n_{r\downarrow} + V_0 \sum_{\langle r, r' \rangle} n_r n_{r'} , 
\end{align}
where $\Delta_0$ is the staggered sublattice potential, $U_A$ and $U_B$ are on-site interactions, and $V_0$ the nearest-neighbor repulsion. 
We consider this two band model at or slightly above the filling of $n=2$ electrons per unit cell.

To controllably describe the effects of interband processes on doped electrons, we will consider situations where the two bands are weakly coupled. 
This is realized in the two following regimes, that we respectively develop in Secs.~\ref{sec:KineticExpansion} and~\ref{sec:InteractionExpansion}: 
\begin{itemize}
\item[--] 
When $t_0=0$, the lower and upper bands of our model are formed by $A$ and $B$ sublattice states respectively. 
A small tunneling amplitude $t_0 \ll \Delta_0$ induces weak hybridization between these sublattices/bands, 
which we treat as a perturbation to derive an effective model for doped electrons. 
\item[--] When the interactions are small compared to the single particle band gap $\Delta_0$, the two bands are weakly coupled by the interaction terms. 
This regime, often studied with standard many-body perturbation theory, offers a stringent test of our kinetic expansion method since their domains of validity overlap when $(t_0,V_0) \ll \Delta_0$. 
\end{itemize}
In both limits, the same essential physics -- attractive interaction induced by 
virtual interband processes, sketched in Fig.~\ref{fig_EffectiveModel}a -- is at play, as highlighted by the consistency between kinetic and interaction based perturbative methods.

\section{Kinetic Energy Expansion} \label{sec:KineticExpansion}

\subsection{Effective dynamics of charge carriers}

\begin{figure}
\centering
\includegraphics[width=0.85\columnwidth]{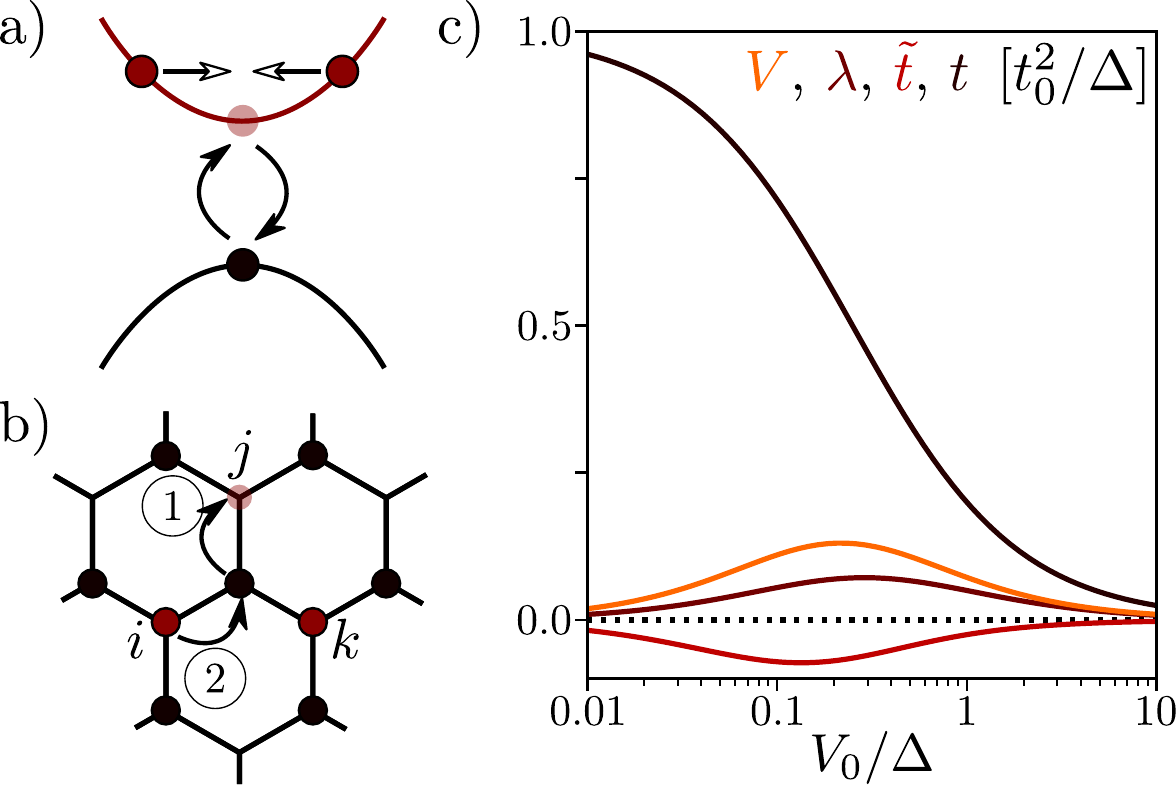}
\caption{
a) Virtual interband transitions involving three electrons in the upper band mediate effective pairing interactions between conduction electrons.
b) Second order process of $\mathcal{H}_f$ corresponding to correlated hopping.  Doubly occupied orbitals forming the band insulator are shown as black dots, doped electrons in red. 
c) The hopping and interaction amplitudes of $\mathcal{H}_f$ as a function of the original lattice parameters $V_0/\Delta$ for $U_B=4V_0$.
} \label{fig_EffectiveModel}
\end{figure}

When $\Delta \equiv  \Delta_0 - U_A$ is much greater than $t_0$, the ground state of Eq.~\ref{eq_original_model} at $n=2$ is an insulator, which is adiabatically connected to the noninteracting band insulator obtained when $U_A=U_B=V_0=0$. 
This state is also adiabatically connected to the $t_0=0$ limit where $A$ sites of the honeycomb lattice are doubly occupied and $B$ sites are empty.
Pauli exclusion principle requires doped electrons above $n=2$ to live on $B$ sites, which form a triangular lattice. 
Our following analysis is based on a perturbative expansion in the tunneling term around the atomic limit with $t_0=0$~\cite{2020arXiv201208528C}.   
Since direct tunneling to doubly occupied $A$ sites is prohibited, the effective Hamiltonian governing the dynamics of the $x \equiv n-2$ doped electrons, denoted as $\mathcal{H}_f$, arises from second-order tunneling processes where the first step necessarily creates a hole on an $A$ site, resulting in intermediate states involving excitons (see Fig.~\ref{fig_EffectiveModel}b).

To account for all such processes, we perform a Schrieffer-Wolff transformation that integrates out high-energy degrees of freedom~\cite{schrieffer1966relation}, and obtain the effective Hamiltonian
\begin{eqnarray}  
\mathcal{H}_f & =& t \sum_{\langle i,j \rangle, \sigma}  ( f_{j,\sigma}^\dagger  f_{i,\sigma} + hc )  + U \sum_{i} n_{i,\uparrow} n_{i,\downarrow} + V \sum_{\langle i,j\rangle} n_i n_j \nonumber \\
&+& \frac{\tilde{t}}{2}  \sum_{\langle i,j \rangle, \sigma} ( f_{j,\sigma}^\dagger  f_{i,\sigma}  + hc)  ( n_i+n_j)  \nonumber \\
& +& \lambda \sum_{ijk\in\triangle,\sigma} \left[ f_{j,\sigma}^\dagger  n_k  f_{i,\sigma}  + P_{ijk} \right]  \label{eq_effectivemodelBfermions}
\end{eqnarray}
where the fermion operators $f_i$ describe electrons on the triangular lattice formed by $B$ sites. 
Besides single-particle hopping ($t$), on-site and nearest-neighbor repulsion ($U, V$), this effective Hamiltonian contains two types of correlated hopping terms: an electron hops between $i$ and $j$ only when either site is occupied by a second electron of opposite spin ($\tilde{t}$), or when their common neighbor $k$ is occupied ($\lambda$). The second type -- hereafter referred to as $\lambda$-hopping -- acts on upper triangles $(ijk \in \triangle)$. 
$P_{ijk}$ denotes all possible permutations of the vertices $ijk$. 
$\mathcal{H}_f$ also contains three-body interactions (see App.~\ref{app:UnitaryTransformation}), but their effect is negligible in the low density limit.

$\mathcal{H}_f$ is exact up to second order in the hybridization parameter $t_0/\Delta$ that governs the perturbative Schrieffer-Wolff transformation, holds at all dopings, and features interactions between doped electrons that are {\it instantaneous} on the time scale set by their kinetic energy $t$. 
The density-density interaction and correlated hopping terms between doped electrons arise because the bare interactions ($U_A, U_B, V_0$) affect the energy of intermediate exciton states. 
For example, the process leading to $\lambda$-hopping is shown in Fig.~\ref{fig_EffectiveModel}b. The presence of a charge at site $k$ in the upper triangle $ijk$ decreases the energy of the intermediate state in the electron tunneling event $f_{j,\sigma}^\dagger f_{i,\sigma}$ from $\Delta + 4V_0$ to $\Delta +3V_0$~\cite{slagle2020charge}.
The hopping amplitude accordingly increases from $t$ to $t+\lambda$, with 
\begin{subequations} \label{eq:CoefficientSchriefferModel} \begin{align}
t & = \frac{t_0^2}{\Delta + 4 V_0} , \\  \lambda & =  \frac{t_0^2}{\Delta + 3 V_0} - \frac{t_0^2}{\Delta + 4 V_0} . 
\end{align} \end{subequations}
Similar considerations yield the other coefficients of $\mathcal{H}_f$ (see App~\ref{app:UnitaryTransformation}), and give $(\tilde{t}, V) \propto t_0^2$ and $U\simeq U_B$ when $U_B \gg t_0^2/\Delta$. 
These parameters are plotted in Fig.~\ref{fig_EffectiveModel}c. 
The bandwidth $W=9t$ largely dominates $\tilde{t}, \lambda, V$, even for rather large bare repulsion $V_0 = 10\Delta$.

Our derivation can be straightforwardly extended to account for longer-range hopping between $B$ orbitals, allowing us to describe materials with a larger bandwidth $W$. 
The effects of longer range interactions can also be included in a similar manner~\cite{2020arXiv201208528C}.

\subsection{Two-particle bound state}

Remarkably, $\mathcal{H}_f$ features an attractive interaction between conduction electrons forming a spin-triplet state. 
This effective attraction surprisingly emerges in the purely repulsive Hubbard model Eq.~\ref{eq_original_model} due to $\lambda$-hopping, as it lowers the energy of two electrons on adjacent sites compared to the case when they are far apart. 
In our model, this attraction competes with and wins over the repulsion $V>0$, which disfavors nearest-neighbor pairing.

\begin{figure}
\centering
\includegraphics[width=\columnwidth]{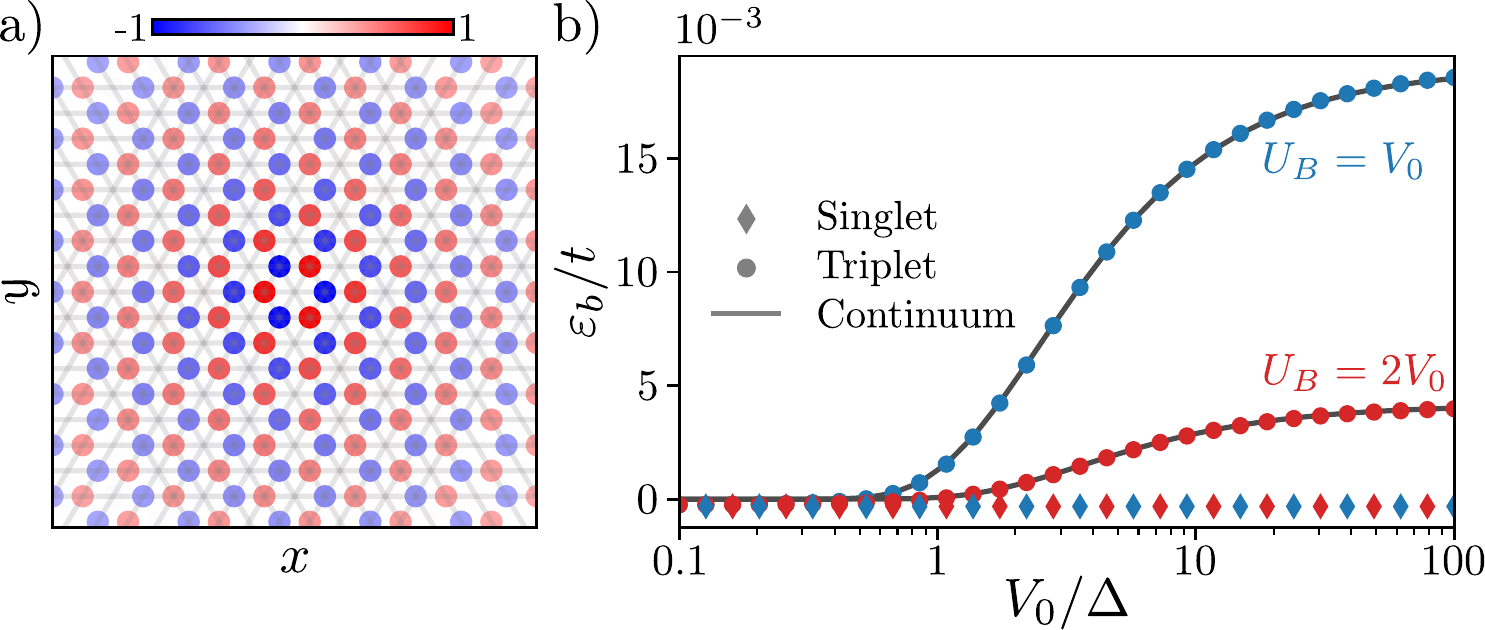}
\caption{ 
a) Amplitude of the triplet bound state wavefunction as a function of the relative distance between the two doped electrons, we choose $U_B = 2 V_0 = 10 \Delta$ for illustration purposes.
b) Binding energy of spin singlet (diamonds) and triplet (circle) pairs, obtained by solving the lattice two-electron problem, or predicted by the continuum model prediction Eq.~\ref{eq_ContinuumBindingEnergy} (gray).
} \label{fig_TwoBodyLattice}
\end{figure}

To demonstrate pairing, we consider the analog of Cooper's problem in a doped band insulator and solve Eq.~\ref{eq_effectivemodelBfermions} for two doped electrons (see App.~\ref{app:TwoBodySolutionLattice}).
Bound states, signaled by a positive energy $\varepsilon_b$, are found in the spin-triplet channel in a very large parameter range. 
In Fig.~\ref{fig_TwoBodyLattice}a, we represent such triplet bound state as a function of the relative position between the two particles. 
This wavefunction exhibits f-wave symmetry, \textit{i.e.} it is symmetric under three-fold rotation and changes sign under any reflection flipping one of the primitive vectors $\vec{a}_j$. 
In Fig.~\ref{fig_TwoBodyLattice}b, we show the binding energy $\varepsilon_b = 6t - E_{2}$, with $E_{2}$ the two-electron ground state energy, as a function of the the original model parameters $U_B, \Delta$ and $V_0$. 
The binding energy reaches the maximum value $\epsilon_b=0.89t$ when the condition $V_0\gg \Delta, U_B$ is satisfied. 
Increasing $U_B$ suppresses the contribution of intermediate states involving doubly occupied $B$ sites, thus leading to a reduced binding energy. 
In the case $U_B = 1.5V_0\gg \Delta$, we have $\varepsilon_b=0.088t$.  
In the limit $U_B \gg V_0$ and $\Delta$, bound state formation requires $V_0>\Delta$. In contrast to triplet pairs, no bound state is found for spin-singlet pairs.

It is important to note that in our model, triplet pairing already occurs at {\it infinitesimal} carrier doping in a band insulator with purely repulsive interaction. It is remarkable that electron pairing by repulsion occurs in such a simple system, without any connection to resonating valence bond or 
quantum spin liquid.

We further show in App.~\ref{app:TwoBodySolutionLattice} that the formation of two-particle bound state is perturbatively stable against the inclusion of longer range part of the Coulomb repulsion. Moreover, at finite doping concentration, Coulomb interaction is dynamically screened by free carriers and effectively becomes short-ranged. Our pairing mechanism relies on  short-range repulsion (such as $V_0$ between adjacent $A$ and $B$ sites) that couples itinerant electrons in the conduction band with core electrons in the filled band. Such interband effects produces attractive pairing interaction, as we have demonstrated rigorously above. Thanks to the beneficial effect of dynamical screening, our interband electronic mechanism for superconductivity can work at finite doping, even when two-particle bound state does not exist \cite{2020arXiv201208528C}.

\subsection{Superconductivity in the dilute limit} 

The existence of spin-triplet bound states is suggestive of superconductivity at finite doping. 
We now study the Hamiltonian $\mathcal{H}_f$ for doped electrons at finite density on the triangular lattice. The single-particle dispersion, given by $\varepsilon_{\bf k} = 2 t \sum_{j=1}^3 \cos ( {\bf k} \cdot {\bf a}_j)$, gives rise to two degenerate minima at $K$ and $K'$ points of the Brillouin zone. 
At sufficiently small density $x \ll 1$, electrons primarily occupy states near these minima. 
Thus, our system at small doping is an interacting two-valley electron liquid.  

We now derive an effective continuum theory that captures the long-wavelength behavior of this liquid. 
Retaining only fermionic modes $ f_{\sigma}(\tau K + {\bf k}) \equiv \psi_{\sigma\tau}({\bf k})$, $\sigma \in \{\uparrow,\downarrow\}$, close to either of the two valleys $\tau \in \{ K, K' \}$, we derive the effective continuum description of the lattice model $\mathcal{H}_f$
\begin{equation} \label{eq_effectiveontinuum_kinetic}
\tilde{\mathcal{H}} = \tilde{\mathcal{H}}_0 + \tilde{\mathcal{H}}_i , \quad \tilde{H}_0 = \int {\rm d} x \sum_{\sigma \tau} \psi_{\sigma\tau}^\dagger \left[ \frac{-\nabla^2}{2m} \right] \psi_{\sigma\tau} ,
\end{equation}
with $m=2/(3ta^2)$ the effective mass at $K$ and $K'$. 
$\tilde{\mathcal{H}}_i$ consists of three symmetry-allowed contact interactions:  
\begin{equation}  \label{eq_continuum_theory}
\tilde{\mathcal{H}}_i = \int {\rm d}x \; g_0 
(\rho_{K\uparrow} \rho_{K\downarrow} + \rho_{K'\uparrow} \rho_{K'\downarrow})  
+ g_1 \rho_{K} \rho_{K'} +  g_2 {\bf s}_K  \cdot {\bf s}_{K'}  
\end{equation}
where $\rho_{\tau} = \psi_{\alpha,\tau}^\dagger \psi_{\alpha,\tau}$ and ${\bf s}_\tau = \psi_{\alpha,\tau}^\dagger \bm{\sigma}_{\alpha \beta} \psi_{\beta,\tau}$ respectively denote the density and spin at valley $\tau$. 
Here, $g_0$ and $g_1$ respectively correspond to intra- and inter-valley repulsion, while $g_2$ describes inter-valley exchange interactions.  

By expressing the microscopic lattice model Eq.~\ref{eq_effectivemodelBfermions} in terms of low-energy modes, we derive the three coupling constants (see App.~\ref{app:ContinuumLimit}) 
\begin{subequations} \label{eq:CouplingConstantFieldTheory} \begin{align}
g_0 & =  (U + 6 V - 6 \lambda - 6 \tilde{t} ) / \mathcal{A} & & > 0 , \\ 
g_1 & = (U +15V - 24 \lambda - 6 \tilde{t}) / (2\mathcal{A})   & & > 0 , \\ 
g_2 & = - 2 (U-3V+12\lambda-6\tilde{t}) / \mathcal{A} & & < 0 , 
\end{align} \end{subequations}
with $\mathcal{A}=2/\sqrt{3} a^2$ the Brillouin zone area.
The emergence of {\it ferromagnetic} inter-valley exchange interaction ($g_2<0$) can be intuitively understood as Hund's rules applied to the valley degree of freedom.

The continuum theory Eq.~\ref{eq_continuum_theory} predicts the existence of spin-triplet valley-singlet bound states in the s-wave channel when 
\begin{equation} \label{eq_negative_coupling_constant}
g = g_1+g_2/4 = 9 (V-2\lambda)/\mathcal{A} < 0 .
\end{equation}
Importantly, these valley-singlet bound states correspond to f-wave electron pairs on the lattice, as they are invariant under threefold rotation $\psi_K \rightarrow e^{i 2\pi/3} \psi_K, \psi_{K'}\rightarrow e^{-i2\pi/3} \psi_{K'}$, but change sign under reflections exchanging the two valleys $\psi_K \leftrightarrow \psi_{K'}$. 
Their binding energy is given by~\cite{levinsen2015strongly}
\begin{equation} \label{eq_ContinuumBindingEnergy}
\varepsilon_b = \left. \Lambda \middle/ \left[ \exp \left( \frac{4\pi}{m|g|} \right) - 1 \right] \right., 
\end{equation}
with $\Lambda$ a UV energy cutoff. 
As shown in Fig.~\ref{fig_TwoBodyLattice}b, this formula almost perfectly reproduces our solution of the two-body problem on the lattice with $\Lambda \simeq 1.75 t$ as the only fitting parameter. 
This proves the validity of our derivation of the continuum theory Eq.~\ref{eq_continuum_theory} from the microscopic lattice model.

\begin{figure}
\centering
\includegraphics[width=\columnwidth]{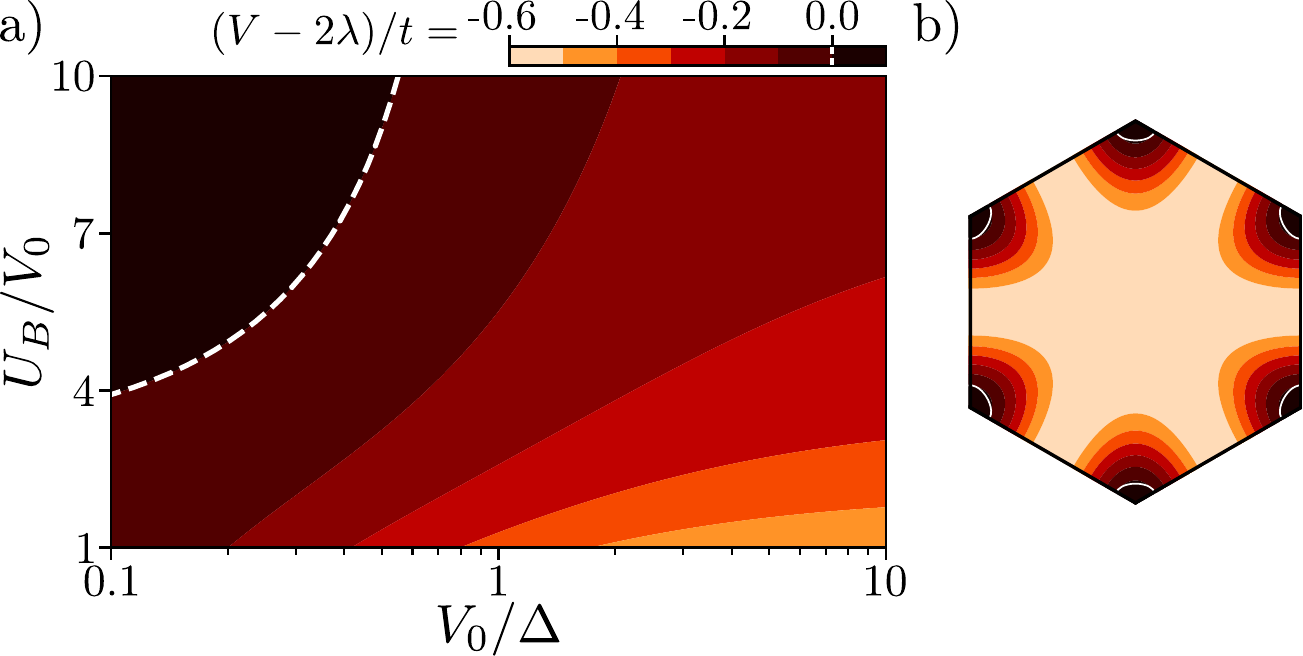}
\caption{ 
a) The coupling strength in the spin-triplet f-wave channel $\propto (V - 2\lambda)$ is negative for most lattice parameters, heralding attractive interactions and superconductivity.
b) Pairing vector amplitude $|\vec{d}_q|^2$ over the Brillouin zone. At low doping, it is isotropic and fully gapped over the disconnected Fermi surface, which is shown as a white line for $x=0.1$. 
} \label{fig_RegionPairing}
\end{figure}

The condition $g<0$ provides an analytical criterion for the appearance of pairing in terms of the original lattice parameters $\Delta_0, U_A, U_B$ and $V_0$. 
Using Eqs.~\ref{eq:CoefficientSchriefferModel}, we observe that this criterion is satisfied in a wide parameter window, which is shown in Fig.~\ref{fig_RegionPairing}a. 
Note that in the presence of on-site repulsion $U_B \neq 0$, pairing occurs only when the nearest-neighbor repulsion $V_0$ exceeds a critical value that depends on $U_B$ and $\Delta$. 
To capture such effect, it is thus essential that our theory of exciton-mediated pairing is non-perturbative in interaction strength.

\subsection{$T_c$ versus doping}

At finite density, the spin-triplet valley-singlet pairing leads to a superconducting state with a vector order parameter 
\begin{eqnarray}
\vec{d} = \langle \psi_{K,\alpha} [\bm{\sigma}  (i\sigma_y)]_{\alpha, \beta}  \psi_{K', \beta} \rangle. 
\end{eqnarray}
This state is fully gapped and isotropic, \textit{i.e.} the order parameter $\vec{d}$ is constant (see Fig.~\ref{fig_RegionPairing}b). 
This property is unusual for triplet superconductors, where the antisymmetry of the Cooper pair wavefunction implies $\vec{d}({\bf k})=-\vec{d}(-{\bf k})$. 
For a singly connected Fermi surface centered at ${\bf k}=0$, the condition necessarily requires strong variations of the $\vec{d}$-vector over the Fermi surface, as exemplified in the $A$ and $B$ phases of $^3$He. 
In stark contrast, in our case the presence of two disconnected Fermi surfaces centered around $K$ and $K'$ enables inter-valley spin-triplet pairing with a constant and opposite $\vec{d}$-vector over the Fermi surface of each valley.

Another important feature of our theory is the non-retarded pairing interaction, which spreads over the entire bandwidth of doped charges. 
This sharply contrasts with electron-phonon superconductors, where the attractive interaction is cut off by the Debye energy, and thus limited to the vicinity of the Fermi surface. 
The absence of energy cutoff changes the expression of the superconducting critical temperature $T_c$. We find that the mean-field $T_c$ depends strongly on the carrier density $x$, despite the constant density of states~\cite{randeria1990superconductivity}
\begin{equation} \label{eq_TcContinuumMeanField}
k_B T^{\rm MF}_c = \frac{e^{\gamma}}{\pi} \sqrt{\varepsilon_F \varepsilon_b} \propto  \sqrt{x W} \exp \left( -\frac{4\pi}{m|g|} \right) .
\end{equation}
where $\varepsilon_F$ is the noninteracting Fermi energy. As a consequence, the gap to $T_c$ ratio exceeds the BCS value of 1.764 by almost a factor of 3~\cite{2020arXiv201208528C}. 
At very low density where $\varepsilon_F<\varepsilon_b$, the BCS mean-field approximation is inadequate. The superconducting state is instead a Bose-Einstein condensate of spin-triplet pairs.

At high density, lattice effects become important. We have performed  mean-field calculations for the lattice model  Eq.~\ref{eq_effectivemodelBfermions} (see App.~\ref{app:MeanFieldSC}). 
In agreement with the previous discussion, we find that the spin-triplet f-wave pairing channel is the leading instability in our model.
The mean-field order parameter takes the form 
\begin{eqnarray}
\vec{d}_{\bf k} =  \vec{d} \sum_j  \sin( {\bf k}\cdot {\bf a}_j)   \equiv s_{\bf k} \vec{d}. 
\end{eqnarray}
Its dependence on the crystal momentum, shown in Fig.~\ref{fig_RegionPairing}b, is determined by the form of interactions. The overall amplitude of $\vec{d}$ is obtained from the spin-triplet f-wave gap equation
\begin{equation} \label{eq_TcInFwave}
\frac{3}{2\lambda - V} = \int \frac{{\rm d}^2 {\bf k}}{\mathcal{A}} \frac{s_{\bf k}^2}{E_{\bf k}} \tanh \left( \frac{E_{\bf k}}{2k_B T} \right) , 
\end{equation}
which always has a solution in the pairing region identified by Eq.~\ref{eq_negative_coupling_constant} (see Fig.~\ref{fig_RegionPairing}a). Here, the quasi-particle energy spectrum
\begin{equation}
E_{\bf k} = \sqrt{(\varepsilon_{\bf k}-\mu)^2 + |\vec{d}_{\bf k}|^2} , 
\end{equation}
with $\mu$ the chemical potential, remains gapped at the Fermi level throughout the transition from the band insulator to the spin-triplet superconductor (recall that $|\vec{d}_{\bf K}|=|\vec{d}_{\bf K'}|\neq 0$).

\begin{figure}
\centering
\includegraphics[width=\columnwidth]{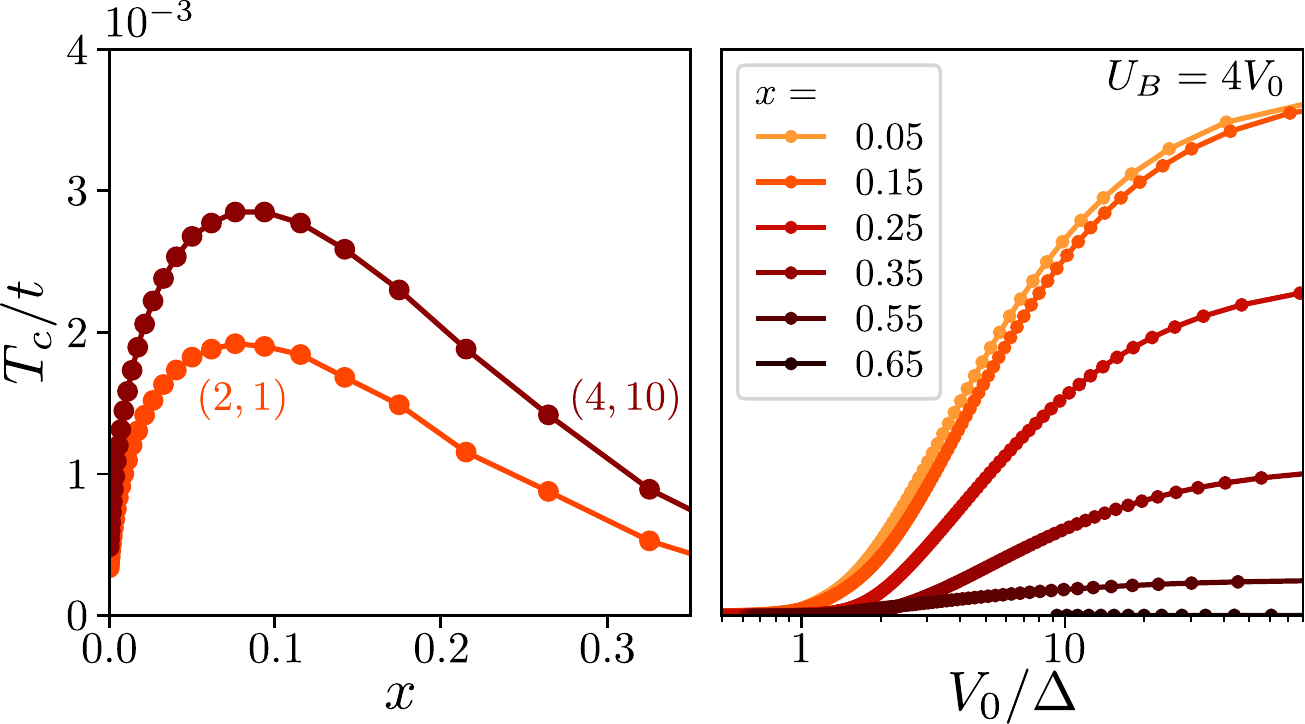}
\caption{ 
Critical temperature $T_c$ (a) as a function of $x$, simulation parameters are indicated as $(U_B/V_0,V_0/\Delta)$, and (b) as a function of $V_0/\Delta$ for $U_B=4V_0$ and several doping concentrations.
}
\label{fig_TCasFctDoping}
\end{figure}

In Fig.~\ref{fig_TCasFctDoping}, we show the critical temperature $T_c$ extracted from Eq.~\ref{eq_TcInFwave} as a function of $x$ and $V_0/\Delta$. At small doping, the f-wave pairing vector $|\vec{d}_{\bf k}|$ is nearly constant around $K$ and $K'$.  
$T_c$ sharply rises with $\sqrt{x}$ dependence, in agreement with the continuum prediction Eq.~\ref{eq_ContinuumBindingEnergy}.
As doping increases, the Fermi surface approaches the $\Gamma-M$ lines, where the order parameter $\vec{d}_{\bf k} \propto s_{\bf k}$ vanishes.
Its small amplitude close to these lines leads to a reduction of $T_c$ for $x>0.1$, as can be seen in Fig.~\ref{fig_TCasFctDoping}a.

Our results for $x \geq 0.5$ shows that $T_c$ almost vanishes in the strongly doped regime where the f-wave superconducting order parameter has nodes, see Fig.~\ref{fig_TCasFctDoping}b. 
In this limit, other studies based on weak-interaction expansions suggest that f-wave and d-wave superconductivity compete~\cite{xiao2016possible,kuroki2010spin}. 
Our strong coupling analysis also finds an attractive d-wave pairing channel in a small region of the parameter space (see App.~\ref{app:MeanFieldSC}). 
While, for the parameters considered, this d-wave amplitude is always subleading, it can win over the f-wave superconductivity when longer range repulsion are included. 
For instance, a next-nearest neighbor $V_2 > 0$ automatically penalizes the f-wave $B-B$ next-nearest neighbor pairing, but can be accommodated by a d-wave superconducting state which features $A-B$ nearest neighbor pairing. 
Similarly, when the sublattice potential $\Delta_0$ decreases, doped charges start to populate $A$ sites, which may again favor a nearest neighbor d-wave order.  
Both effects are further enhanced near Van Hove doping, where nodeless d-wave states are provably more stable than nodal f-wave order in the limit $\Delta_0=0$~\cite{nandkishore2014superconductivity}.

\section{Interaction expansion} \label{sec:InteractionExpansion}

In this section, we show that the three-particle mechanism for superconductivity described in Sec.~\ref{sec:KineticExpansion} is not restricted to the atomic limit $t_0 \ll \Delta$, but extends at any value of the tunneling parameter. 
To preserve analytical control of interband effects, we work in the small interaction limit, such that interband hybridization remains small. 
In this regime, we find evidence of attractive interaction between conduction electrons.  using a similar method than in Sec.~\ref{sec:KineticExpansion}. 
We follow the methodology of Sec.~\ref{sec:KineticExpansion}, \textit{i.e.} we first derive an effective model for doped electrons with an unitary transformation, and then study the obtained model for small doping concentrations.

\subsection{Unitary transformation}

We start from the Hamiltonian Eq.~\ref{eq_original_model} written in momentum space as $\mathcal{H} = \mathcal{H}_0 + \mathcal{V}$ with
\begin{eqnarray} \label{eq_generic_hamiltonian_momentum}
\mathcal{H}_0 = \sum_{1} \varepsilon_{1} c_{1}^\dagger c_{1} , \quad \mathcal{V} = \frac{1}{N_s} \sum_{1234} V_{43}^{21} \delta_{43}^{21}  c_4^\dagger c_3^\dagger c_2 c_1 , 
\end{eqnarray}
where we have used a generalized index $i = (\vec{k}_i, b_i, \sigma_i)$ gathering the momentum $\vec{k}_i$, band $b_i = \pm$, and spin $\sigma_i$ labels. 
Our goal is to derive an effective Hamiltonian for the upper band assuming the lower one fully filled. 
We do so by eliminating the direct interband mixing interaction terms in the Hamiltonian with the help of a unitary transformation.

This transformation is carried out explicitly in App.~\ref{app:WeakInteraction}, where we find the leading corrections to the dispersion relation $\varepsilon$ and the scattering vertex $V$. 
The former is akin to the Hartree-Fock correction of standard many-body perturbation theory:
\begin{equation} \label{eq:WeakIntCorrectionDispRel} \begin{split}
\delta \varepsilon_{\vec{k},+} & = \frac{1}{N_s} \sum_{\vec{q},\sigma} \left[ V_{(\vec{q},-,\sigma)(\vec{k},+,\uparrow)}^{(\vec{k},+,\uparrow)(\vec{q},-,\sigma)} +  V_{(\vec{k},+,\uparrow)(\vec{q},-,\sigma)}^{(\vec{q},-,\sigma)(\vec{k},+,\uparrow)} \right] \\ 
& - \frac{1}{N_s} \sum_{\vec{q}} \left[ V_{(\vec{k},+,\uparrow)(\vec{q},-,\uparrow)}^{(\vec{k},+,\uparrow)(\vec{q},-,\uparrow)} + V_{(\vec{q},-,\uparrow)(\vec{k},+,\uparrow)}^{(\vec{q},-,\uparrow)(\vec{k},+,\uparrow)} \right] ,
\end{split} \end{equation}
where we have used the spin-independence of $V$, and choose $\uparrow$ as a preferred spin index. 
This small correction does not change the position of the band minima, which remain degenerate at the $K$ and $K'$ points. 
Expanding around these minima in the limit $t_0 \ll \Delta_0$ gives $(\varepsilon + \delta\varepsilon)_{\tau K + \vec{k},+} \simeq |\vec{k}|^2/(2m^*)$ with
\begin{equation} \label{eq:WeakIntEffMass}
\frac{1}{m^*} = \frac{3t_0^2 a^2}{2 \Delta_0^2} \left[ \Delta_0 + U_A - 4 V_0 \right] , 
\end{equation}
which agrees with the result of Eq.~\ref{eq_effectiveontinuum_kinetic} obtained with the kinetic expansion provided interactions are small compared to $\Delta_0$. 
This offers an important consistency test between the two methods in their overlapping regimes of validity.

\subsection{Attraction in dilute limit}

The expressions for the corrections to the scattering vertex $\delta V$ being quite involved (see App.~\ref{app:WeakInteraction}), we simply support here the emergence of attractive interactions in the $f$-wave scattering channel by computing the effective interaction strength in the spin-triplet valley-singlet channel, \textit{i.e.} between equal-spin electrons living in opposite valleys. 
Writing $\tilde{V} = V + \delta V$, this coefficient can be expressed as 
\begin{equation} \begin{split}
U_0 = & \tilde{V}_{(K+)(K'+)}^{(K'+)(K+)}  + \tilde{V}_{(K'+)(K+)}^{(K+)(K'+)} \\ 
& - \tilde{V}_{(K+)(K'+)}^{(K+)(K'+)} - \tilde{V}_{(K'+)(K+)}^{(K'+)(K+)} , 
\end{split} \end{equation}
where all spin component are equal. 
This coefficient greatly simplifies as the Bloch states at the $K$ and $K'$ point in the upper band are localized on $B$ sites, $\Psi_{K/K',+}^A = 0$. 
Using this simplification, we end up with
\begin{equation}
U_0 = \frac{12 t_0^2 V_0(U_B-3V_0)}{N_s} \sum_{\vec{q}} \frac{|f(\vec{q})|^2}{(2\varepsilon_{\vec{q},+})^3} , 
\end{equation}
which also agrees with the our results of Sec.~\ref{sec:KineticExpansion} in the limit where both methods are analytically controlled (see App.~\ref{app:WeakInteraction}). 
The presence of an effective attraction $U_0<0$ in the f-wave channel when $U_B<3V_0$ proves that the results of Fig.~\ref{fig_RegionPairing} hold for non-perturbative tunneling amplitudes $t_0$.

Although the three-particle mechanism for superconductivity introduced in this article is more easily pictured near the atomic limit (Sec.~\ref{sec:KineticExpansion}), it can also apply to band insulators where the conduction band's width is substantial compared to the single particle gap. 
Strong of this generalization, we now compare the predictions of our theory to experimental data on two dilute superconductors.

\section{Application to dilute superconductors}

Let us recapitulate our results so far. 
Using the specific model Eq.~\ref{eq_original_model} as an example, we have demonstrated a general mechanism of spin-triplet superconductivity in doped band insulators with strong repulsive interaction. 
Focusing on the low density regime, we have derived a universal continuum model for the two-valley Fermi liquid. 
When either tunneling or interactions are small compared to the $n=2$ insulator gap, we have shown that virtual interband transitions can produce a sufficiently strong intervalley ferromagnetic interaction that leads to spin-triplet pairing. 
A remarkable consequence of our theory is that spin-triplet superconductivity occurs at infinitesimal doping, with a sharp increase $T_c \sim \sqrt{x}$. 
Based on this theory, we hereby propose that the recently discovered dilute superconductors ZrNCl WTe$_2$ are spin-triplet.

\subsection{ZrNCl}

ZrNCl shares many common features with our toy model Eq.~\ref{eq_original_model}. 
At the single particle level, it is a band insulator with a large gap $\Delta \sim  2.5$eV~\cite{weht1999electron,hase1999electronic}, where a single band with quadratic minima at the $K$ and $K'$ points is relevant to describe the physics at low carrier density~\cite{heid2005ab,yin2013correlation}.
Upon electron doping, two-dimensional superconductivity arises in the ZrN planes~\cite{tou2000evidence,tou2005upper}, even for extremely dilute samples $x=0.0038$~\cite{nakagawa2020gate}.
The critical temperatures ranging from $\SI{11.5}{\kelvin}$ to $\SI{19}{\kelvin}$ is hard to explain by electron-phonon mechanism, which makes ZrNCl an unconventional superconductor.
A compelling argument against the phonon-mechanism of pairing is the simultaneous enhancement of $T_c$~\cite{takano2008interlayer} and reduction of electron–phonon interactions, probed by Raman scattering~\cite{kitora2007probing}, when the doping decreases from 0.2 to 0.05.

We now highlight that our theory consistently captures the available experimental results on the superconducting state in ZrNCl~\cite{kasahara2015unconventional}, and propose that this material be regarded as a legitimate candidate for spin-triplet pairing.

First, we find pairing at infinitesimal doping above the band insulator, and predict a smooth crossover from a BEC of pairs to a BCS regime as the Fermi energy increases~\cite{2020arXiv201208528C}, which has recently been observed in extremely high quality ZrNCl samples~\cite{nakagawa2020gate}.
On the BCS side of this crossover, we expect a gap to critical temperature ratio $2\Delta/k_BT_c > 3.5$~\cite{parish2015bcs,gor1961contribution,chubukov2016superconductivity}. 
This is because the induced pairing interaction in Eq.~\ref{eq_effectivemodelBfermions} is instantaneous on the time scale of the inverse bandwidth, so that all carriers in the narrow band contribute to the superconducting gap. 
This contrasts with phonon-mediated retarded attractions, which only spread over a Debye width near the Fermi level and lead the universal ratio of 3.5. 
The gap to $T_c$ ratio measured by scanning tunneling spectroscopy in ZrNCl is about ten~\cite{ekino2013superconducting}, a sign of strong coupling superconductivity and non-retarded pairing interactions which are well captured by our model.

Second, our theory also allows a fine understanding of the critical temperature observed in ZrNCl. 
Using the DFT results for the lattice constant $a=\SI{3.663}{\angstrom}$ and the effective mass at the $K$ point $m=0.580 m_e$~\cite{tanaka2015minimal,yun2017two}, we estimate $t \sim \SI{0.65}{\electronvolt}$. 
With the parameters $U_B = 4V_0 = 4 \Delta$ of Fig.~\ref{fig_TCasFctDoping}, we obtain a critical temperature  $T_c \sim 0.002 t \simeq \SI{15}{\kelvin}$ for $9\%$ doping, which lies very close to the experimentally measured value.  

Our theory also successfully captures the non-monotonic dependence of $T_c$ on doping observed in Refs.~\cite{nakagawa2020gate}. 
The original increase of $T_c$ following from the BEC/BCS physics, while the subsequent decay of $T_c$  when $x= 5\% \to 20\%$~\cite{takano2008interlayer} is interpreted by the larger modulation of the form factor $s_q$ near $\Gamma-M$ lines. 
For the parameters of Fig.~\ref{fig_TCasFctDoping}, $T_c$ at $x=0.2$ equals $2/3$ of its maximum value, which almost quantitatively agrees with the measured value of $40\%$~\cite{nakagawa2020gate}. 

Finally, specific heat measurements point towards a change from an almost isotropic to a highly anisotropic order parameter upon increasing the doping level~\cite{kasahara2009enhancement,taguchi2005specific}, which is consistent with an f-wave superconducting state. 
The unconventional pairing symmetry is further substantiated by the lack of coherence peak in the NMR signal of Ref.~\cite{kotegawa2014strong}.


\subsection{WTe$_2$}

Monolayer WTe$_2$ is a topological insulator with spin-helical edge states~\cite{qian2014quantum,fei2017edge,tang2017quantum,wu2018observation,lau2019influence}. 
Recently, two independent groups discovered a transition from insulating to superconducting state in monolayer WTe$_2$ under electron doping via electrostatic gating~\cite{fatemi2018electrically,sajadi2018gate}. 
Another separate work observed superconductivity in epitaxial thin film of WTe$_2$~\cite{asaba2018magnetic}. 
The origin of superconductivity and the nature of the superconducting state are unknown and have attracted considerable interest~\cite{hsu2020inversion,xie2020spin}.  

Our picture is that superconductivity in electron doped WTe$_2$ is driven by excitonic effects and exhibits spin-triplet pairing. 
Indeed, a recent experiment on insulating WTe$_2$ reported evidence of strong excitonic effects which significantly enhance the single-particle gap~\cite{jia2020evidence}. 
While the electronic structure of WTe$_2$ is far more complicated than our model used to illustrate the exciton pairing mechanism, they both feature two valleys in the conduction band. 
Therefore, electron doped WTe$_2$ at low density is a two valley Fermi liquid described by our continuum theory Eq.~\ref{eq_continuum_theory}. 
If excitons in WTe$_2$ mediate strong enough inter-valley ferromagnetic exchange, our theory should capture the essential physics of its superconducting state.

\begin{figure}
\centering
\includegraphics[width=\columnwidth]{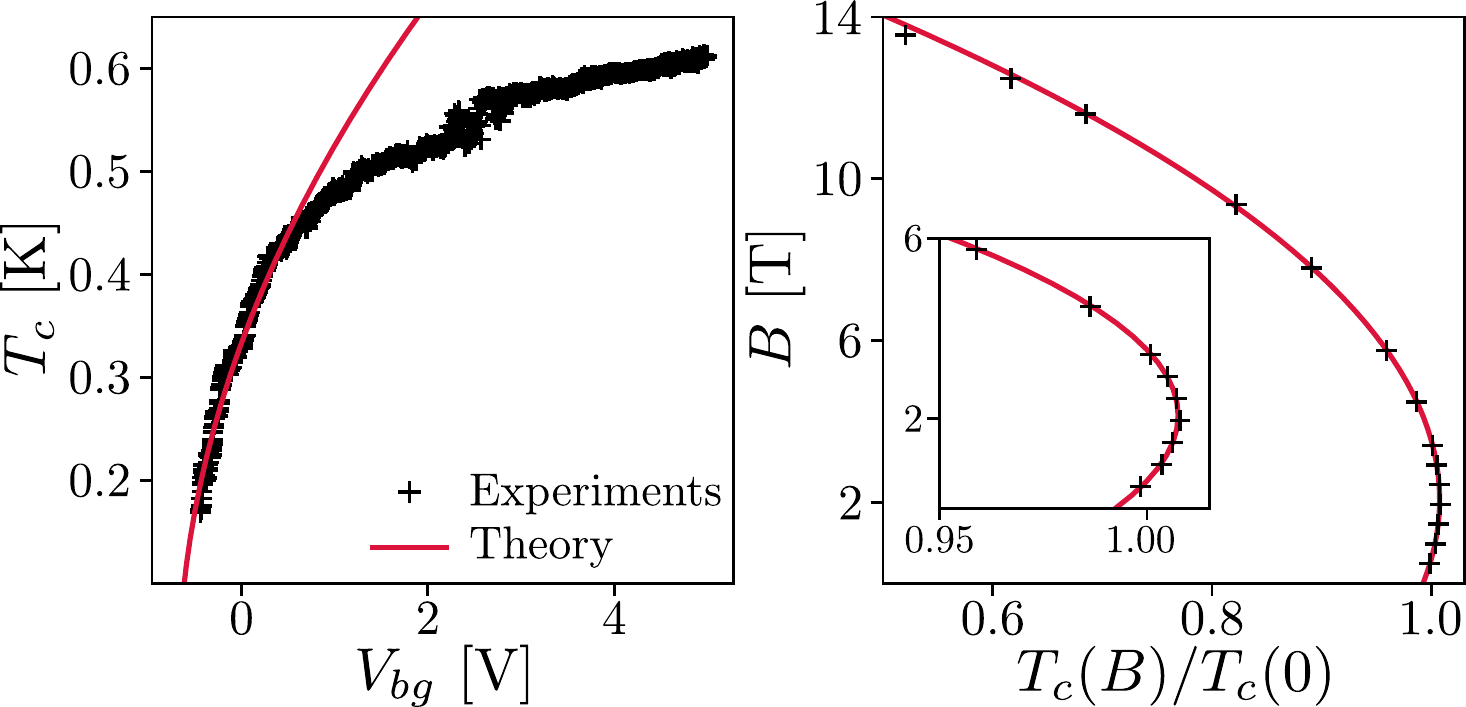}
\caption{ 
The doping dependence of $T_c$, measured in~\cite{fatemi2018electrically}, is well captured by our low-density prediction Eq.~\ref{eq_TcContinuumMeanField}.
b) Perfect agreement is found between the critical temperature measured in Ref.~\cite{asaba2018magnetic} under an in-plane magnetic field $B$ and the prediction Eq.~\ref{eq_cuspTcZeeman}. 
}
\label{fig_ComparisonExperiments}
\end{figure}

This motivates a thorough comparison between the expected consequences of spin-triplet superconductivity to experimental findings on WTe$_2$.   
First, differential resistance measurements~\cite{fatemi2018electrically} shows that upon doping, an insulating resistance peak transforms directly into to a superconducting resistance dip, consistent with our picture of an insulator-superconductor transition. 
The experimentally observed sharp increase of $T_c$ with doping~\cite{fatemi2018electrically} agrees remarkably well with the prediction $T_c \sim \sqrt{x}$ in the low density regime, as shown in Fig.~\ref{fig_ComparisonExperiments}. 
Another prediction of our theory is that single-particle gap does not close across the doping induced insulator-superconductor transition, which can be tested in future tunneling measurements.

The scenario of spin-triplet superconductivity in WTe$_2$ is supported by the observation of an in-plane critical field much larger than the Pauli limit in both monolayers and thin films~\cite{fatemi2018electrically,asaba2018magnetic,sajadi2018gate}. 
The most significant experimental evidence of spin-triplet superconductivity is the initial \emph{increase} of $T_c$ upon application of an in-plane magnetic field~\cite{fatemi2018electrically,asaba2018magnetic}. 
This behavior is incompatible with s-wave superconductivity (even after considering the effect of spin-orbit interaction~\cite{xie2020spin}). On the contrary, the enhancement of $T_c$ by magnetic field follows naturally from our equal-spin triplet superconducting state in two-valley systems. 
This can be seen with the Landau free energy:  
\begin{equation} \label{eq:GL_Basic}
F = \alpha (\vec{d} \cdot \vec{d}^*) + \mu \vec{B} \cdot (i \vec{d} \times \vec{d}^*) + \eta  |\vec{B} \cdot \vec{d}|^2 + \chi B^2 (\vec{d} \cdot \vec{d}^*) ,
\end{equation}
with $\alpha = \kappa(T - T_c(B=0))$ and $\kappa, \mu, \eta, \chi > 0$  near $T_c$. 
Importantly, the magnetic field $\vec{B}$ results in a linear Zeeman shift for pairs of total spin $\vec{S} = i (\vec{d} \times \vec{d}^*)$. 
The $\eta$ term  describes the preferred equal-spin pairing with ${\vec d} \perp \bf B$, \textit{i.e.} the spins are aligned or anti-aligned with the field. 
The last term accounts for orbital effect of the in-plane $B$ field, which causes pair breaking.

From the Landau free energy, we obtain
\begin{equation} \label{eq_cuspTcZeeman}
\Delta T_c^B  =  \mu B - \chi B^2  , 
\end{equation}
with $\Delta T_c^B = T_c(B) - T_c(0)$. 
Due to the Zeeman effect on triplet pairs, $T_c$ increases {\it linearly} with $B$ at small field. 
This  non-analytic dependence on $B$ is a consequence of the degeneracy of triplet superconducting states associated with spin degrees of freedom at zero field. 
At such large $B$, orbital effect dominates and reduces $T_c$. As shown in Fig.~\ref{fig_ComparisonExperiments}, the above theoretical curve $T_c(B)$ fits excellently with experimental data.

The presence of spin-orbit coupling in WTe$_2$ is expected to lift the degeneracy between different triplet states. 
Due to a plane mirror symmetry perpendicular to the $a$ crystallographic axis, an additional term $F_{\rm SOC} = 2 \gamma |d_x|^2$ is allowed after including the spin-orbit effect. 
If $\vec{B} \parallel b$, the linear increase of $T_c$ is rounded off by $\gamma$ according to
\begin{equation} 
\Delta T_c^B = \sqrt{(\mu B)^2 + \gamma^2} - |\gamma| - \chi B^2 . \label{eq_TcB2}
\end{equation}
For small spin-orbit coupling $\gamma < \mu^2/2\chi$, the Zeeman term dominates, which still enables the enhancement of $T_c$ for $\vec{d} \propto (\Delta T_c^B + \chi B^2,0 , i\mu B)^T$. 
For larger $\gamma$, $T_c(B)$ decreases monotonously with field. When $\vec{B} \parallel a$, we should distinguish two cases. First if $\gamma >0$, both spin-orbit and magnetic field favor the vector $\vec{d} \propto (0,1,i)^T$, and we recover Eq.~\ref{eq_cuspTcZeeman}. 
Then for $\gamma<0$,  the original spin degeneracy is fully lifted and we observe a competition between two different triplet states with $\vec{d} \parallel a$ and $\vec{d} \perp a$, respectively favored at low and high field. 
This leads to
\begin{equation} \label{eq_TcB3}
\Delta T_c^B =  \begin{cases}  -(\eta+\chi)B^2 & \text{if: } B < B^* \\ \mu B - |\gamma| - \chi B^2 & \text{if: } B > B^*
\end{cases} ,
\end{equation}
which exhibits a kink at $B^* \simeq |\gamma|/\mu$ where the system undergo a first-order phase transition. 
We hope more measurements of $T_c$ as a function of both in-plane magnetic field strength and direction can be performed to establish the highly unusual behaviour predicted by Eqs.~\ref{eq_cuspTcZeeman} and~\ref{eq_TcB3}, which we regard as an unambiguous evidence for triplet superconductivity.

\section{Acknowledgement}

We thank Kin Fai Mak, Lu Li, Valla Fatemi, Pablo Jarillo-Herrero, Sanfeng Wu, Ali Yazdani, Atac Immomglu and Yoshi Iwasa for valuable discussions. We thank Tomoya Asaba and Valla Fatemi for providing raw experimental data for use in Fig.~\ref{fig_ComparisonExperiments}. 

This work was supported by DOE Office of Basic Energy Sciences, Division of Materials Sciences and Engineering under Award DE-SC0018945. 
VC gratefully acknowledges support from the MathWorks fellowship. 
LF was supported in part by a Simons Investigator Award from the Simons Foundation.

\bibliography{Biblio_TripletSC}

\onecolumngrid
\newpage
\begin{center}
\textbf{\large Supplemental material for ``Spin-triplet superconductivity from excitonic effect in doped insulators"} \\[10pt] 
Valentin  Cr\'epel, Liang  Fu \\
\textit{Department of Physics, Massachusetts Institute of Technology, Cambridge, Massachusetts 02139, USA}
\end{center}
\vspace{20pt}
\twocolumngrid

\renewcommand{\thefigure}{S\arabic{figure}} 
\renewcommand{\thesection}{S\arabic{section}} 
\renewcommand{\theequation}{E\arabic{equation}}
\setcounter{figure}{0}
\setcounter{section}{0}
\setcounter{equation}{0}

\section{Effective model in the atomic limit} \label{app:UnitaryTransformation}

\subsection{General framework}

In this appendix, we derive the effective Hamiltonian Eq.~\ref{eq_effectivemodelBfermions} from the original model Eq.~\ref{eq_original_model} by treating the tunneling terms 
\begin{equation}
\mathcal{H}_t = t_0 \sum_{\langle r, r' \rangle, \sigma} \left( c_{r,\sigma}^\dagger c_{r',\sigma} + hc \right) ,
\end{equation}
as a perturbation of the 'classical' part~\cite{slagle2020charge}
\begin{equation}
\mathcal{H}_0 = \sum_{r \in B} \frac{U_B}{2} n_r (n_r-1) + \Delta n_r + \sum_{\langle r, r' \rangle} V_0 n_r n_{r'} ,
\end{equation}
thanks to the large staggered potential $\Delta \gg t_0$. To that purpose, we apply the unitary transformation $\mathcal{H}' = e^{iS} H e^{-iS}$, with $S$ Hermitian and satisfying~\cite{schrieffer1966relation}
\begin{equation}
[\mathcal{H}_0, iS] =  \mathcal{H}_t .
\end{equation}
It leads to the following approximation of the Hamiltonian
\begin{equation} \label{appeq:Schrieffer_SecondOrderGeneric}
\mathcal{H}' = \mathcal{H}_0 + \frac{1}{2} \left[ iS, \mathcal{H}_t \right] + \mathcal{O} ( \mathcal{H}_t S^2)  ,
\end{equation} 
obtained with the Baker-Campbell-Haussdorf formula. To find an explicit representation of $S$, we decompose the tunneling Hamiltonian as
\begin{equation}
\mathcal{H}_t = \sum_{d=\pm 1} \sum_{v=-5}^5 \sum_{u = -1}^1  T_{d,v,u}  ,
\end{equation}
where $T_{d,v,u}$ gathers all tunneling events that change the number of occupied $A$ sites by $d$, the number of nearest neighbor pairs by $v$ and the number of doubly occupied $B$-sites by $u$. In terms of these operators, we find
\begin{equation}
S = -i \sum_{d,v,u} \frac{T_{d,v,u}}{d \Delta + v V_0 + u U_B} .  
\end{equation}
Plugging this expression in Eq.~\ref{appeq:Schrieffer_SecondOrderGeneric}, we obtain
\begin{equation} \label{appeq:Schrieffer_SecondOrderCommutator}
\mathcal{H}' = \mathcal{H}_0 + \frac{1}{2} \sum_{\substack{ d, v, u \\ d', v', u'}} \frac{[T_{d',v',u'}, T_{d,v,u}]}{d' \Delta + v' V_0 + u' U_B} ,
\end{equation}
which is valid up to $\mathcal{O} \left( t_0^3 / \Delta^2 \right)$ corrections.

\subsection{Projection}

The ground state of $\mathcal{H}_0$ for two electrons per unit cell has a singlet on all $A$-sites and $B$-sites completely empty. Due to Pauli exclusion principle, the $x=n-2$ doped electrons above this insulating state are placed at the $B$-sites. They have an energy per particle $E_f = \Delta + 6V_0$. This low energy manifold, named $f$-band, hybridizes with local excitation having a hole on a $A$-site due to the tunneling part $\mathcal{H}_t$. Such local excitation are separated from the low-energy band by an energy of at least $\Delta$. They are only virtually occupied due to the small ratio $t_0 \ll \Delta$, and their effects on the $f$-electrons' dynamics can be obtained with Eq.~\ref{appeq:Schrieffer_SecondOrderCommutator}.

Projecting $\mathcal{H}'$ onto the $f$-band requires to have $d'=-d$ and $v'=-v$ in Eq.~\ref{appeq:Schrieffer_SecondOrderCommutator}. Furthermore, the first operator acting on the $f$-band should move an electron from an $A$ to a $B$ site, \textit{i.e.} the rightmost $T_{d,v,u}$ must have $d = 1$ and $v, u \geq 0$. This gives 
\begin{align} 
\mathcal{H}' \simeq \mathcal{H}_0 - \frac{1}{2} & \sum_{v,u, u'\geq 0}  T_{-1,-u',-v} T_{1,u,v} \\ &
\times \left[ \frac{1}{\Delta + v V_0 + u U_B} + \frac{1}{\Delta + v V_0 + u' U_B} \right]  . \nonumber
\end{align}
This Hamiltonian can be recast as a tight binding Hamiltonian for the $f$-electrons on the triangular lattice, with density-assisted hopping and local interactions:
\begin{equation} \begin{split}
\mathcal{H}'  = & U_B \sum_{i \in B} \frac{n_i (n_i-1)}{2} + \sum_{ijk \in \triangle} \left[ \tilde{V}_{ij,k} + \circlearrowleft_{ijk} \right] \\
& + \sum_{ijk \in \triangle, \sigma} \left[ f_{j,\sigma}^\dagger \tilde{T}_{ij,k} f_{i,\sigma} + P_{ijk} \right] , 
\end{split} \end{equation}
where sums run over upper triangles with vertices $ijk$, while $\circlearrowleft_{ijk}$ and $P_{ijk}$ respectively denote cyclic and all permutations of $ijk$. The density dependent interaction and tunneling operators read
\begin{align}
\tilde{V}_{ij,k} & = - \frac{t_0^2(2-n_k)}{\Delta + (5-n_\triangle) V_0 + n_k U_B} , \\
\tilde{T}_{ij,k} & = \frac{t_{\Delta, i} + t_{\Delta, j}}{2} , \quad t_{\Delta, \ell} = \frac{t_0^2}{\Delta + (4-n_\triangle) V_0 + n_\ell U_B} , \nonumber 
\end{align}
with $n_\triangle =n_i+n_j+n_k$, which are symmetric under the exchange $i \leftrightarrow j$. 
These interaction coefficients and density-dependent tunneling amplitudes can be expressed in terms of sum and product of density operators. 
We now simplify their expression in two particular cases. 

\begin{widetext}
\subsection{Dilute limit}

At small doping concentration, we can discard states with more than two fermions on the same triangle, which only appear with negligible probability. The tunneling coefficient $\tilde{T}_{ij,k}$ is thus restricted to cases where $(n_i+n_j=0,n_k=0)$, $(n_i+n_j=1,n_k=0)$ or $(n_i+n_j=0,n_k=1)$. Projecting on these configurations, we find the equivalent representation
\begin{equation}
\tilde{T}_{ij,k} = T_{0,0} + n_k [T_{0,1}-T_{0,0}] + (n_i+n_j) [T_{1,0}-T_{0,0}] .
\end{equation}
Summing over all possible triangle, we can rewrite
\begin{equation}
\sum_{ijk \in \triangle, \sigma} \left[ f_{j,\sigma}^\dagger \tilde{T}_{ij,k} f_{i,\sigma} + P_{ijk} \right] = \sum_{\langle i,j \rangle, \sigma} \left[ f_{j,\sigma}^\dagger \left[ t + \tilde{t} \frac{n_i+n_j}{2} \right] f_{i,\sigma} + hc \right]  + \lambda \sum_{ijk\in\triangle,\sigma} \left[ f_{j,\sigma}^\dagger  n_k  f_{i,\sigma}  + P_{ijk} \right] ,
\end{equation}
where we have introduced the coefficients
\begin{equation}
t  = \frac{t_0^2}{\Delta + 4V} , \quad 
\tilde{t}  = \frac{t_0^2}{\Delta + 3 V_0 + U_B} + \frac{t_0^2}{\Delta + 3 V_0}  - \frac{2 t_0^2}{\Delta + 4 V_0}  , \quad 
\lambda  = \frac{t_0^2}{\Delta + 3 V_0} - \frac{t_0^2}{\Delta + 4 V_0}.
\end{equation}
Similarly, we can project the interaction terms on configurations having $n_k, (n_i+n_j), (n_i+n_j+n_k) \leq 2$:
\begin{align}
\tilde{V}_{ij,k} & = \tilde{V}_{ij,0} + n_k [\tilde{V}_{ij,1}-\tilde{V}_{ij,0}] + \frac{n_k(n_k-1)}{2} [\tilde{V}_{0,0}-2\tilde{V}_{0,1}] \\
& = \tilde{V}_{0,0} + (n_i+n_j)[\tilde{V}_{1,0}-\tilde{V}_{0,0}]  + \frac{n_i(n_i-1) + n_j (n_j-1) + 2n_in_j}{2} [\tilde{V}_{2,0}-2\tilde{V}_{1,0}+\tilde{V}_{0,0}] \\
& \quad + n_k [ \tilde{V}_{0,1}-\tilde{V}_{0,0} + (n_i+n_j) ( \tilde{V}_{1,1}-\tilde{V}_{1,0} - \tilde{V}_{0,1} + \tilde{V}_{0,0} ) ] + \frac{n_k(n_k-1)}{2} [\tilde{V}_{0,0}-2\tilde{V}_{0,1}] .
\end{align}
Up to a global constant and a shift of chemical potential, this expansion leads to
\begin{equation}
U_B \sum_{i \in B} \frac{n_i (n_i-1)}{2} + \sum_{ijk \in \triangle} \left[ \tilde{V}_{ij,k} + \circlearrowleft_{ijk} \right] = U \sum_{i} \frac{n_i(n_i-1)}{2} + V \sum_{\langle i,j\rangle}  n_i n_j ,
\end{equation}
where the coefficients read
\begin{equation}
U_f = U + 2 \tilde{V}_{2,0} - 4\tilde{V}_{1,0} - 2\tilde{V}_{0,1} + 3\tilde{V}_{0,0} , \quad V_f = \tilde{V}_{2,0} + 2 \tilde{V}_{1,1} - 4\tilde{V}_{1,0} - 2\tilde{V}_{0,1} + 3\tilde{V}_{0,0} .
\end{equation}
In terms of the lattice parameters $t_0$, $\Delta$, $V_0$ and $U_B$, we find them equal to
\begin{align}
V & = \frac{4t_0^2 V_0 (2V_0+\Delta)}{(\Delta+3V_0) (\Delta+4V_0) (\Delta+5V_0)} - \frac{ 2 t_0^2 V_0}{(\Delta+3V_0+U_B)(\Delta+4V_0+U_B)} , \\
U & = U_B -\frac{2t_0^2 (16V_0^2+7 V_0 \Delta + \Delta^2)}{(\Delta+3V_0) (\Delta+4V_0) (\Delta+5V_0)} + \frac{2t_0^2}{\Delta+4V_0+U_B} .
\end{align}
Gathering the various terms, we obtain Eq.~\ref{eq_effectivemodelBfermions} of the main text.

\subsection{Large $U$ limit}

Assuming $U_B \gg t_0$, we can project the effective Hamiltonian to the $f$-band with no double occupancy. Restricting to $n_i \leq 1$ in the above equations yields the effective Hamiltonian 
\begin{equation}  \label{appeq:HamiltonianLargeU}
\mathcal{H}' = \sum_{\langle i, j \rangle, \sigma} t \left( f_{i,\sigma}^\dagger f_{j,\sigma} + hc \right) + V n_i n_j + \sum_{(ijk) \in \triangle} \lambda  \left( \sum_\sigma f_{i,\sigma}^\dagger  n_k f_{j,\sigma} + P_{ijk} \right) + U_3 n_i n_j n_k  , 
\end{equation}
where the coefficients $t$, $\lambda$ and $V$ have the same form as above. The three-body interaction terms read
\begin{equation} 
U_3 = \frac{12 t_0^2 V_0^2}{(\Delta+3V_0)(\Delta+4V_0)(\Delta+5V_0)} - \frac{6 t_0^2 V_0^2}{(\Delta+2V_0+U_B)(\Delta+3V_0+U_B)(\Delta+4V_0+U_B)} .
\end{equation}

\end{widetext}

\section{Two-particle lattice calculation} \label{app:TwoBodySolutionLattice}

In this appendix, we solve the effective model Eq.~\ref{eq_effectivemodelBfermions} for two particles -- the analog of Cooper's problem on the lattice. To do so, we separate the center of mass momentum $K$ from the relative motion with the introduction of the states 
\begin{align}
\ket{\varphi_0 (K, r)} & = \frac{1}{\sqrt{N_s}} \sum_R e^{i (K \cdot R)} ( f_{R,\uparrow}^\dagger f_{R+r,\downarrow}^\dagger + f_{R+r,\uparrow}^\dagger f_{R,\downarrow}^\dagger ) , \nonumber \\ 
\ket{\varphi_1 (K, r)} & = \frac{1}{\sqrt{N_s}} \sum_R e^{i (K \cdot R)} f_{R,\uparrow}^\dagger f_{R+r,\uparrow}^\dagger ,
\end{align}
where the subscripts denotes the total spin $S$ of the state (singlet $S=0$ or triplet $S=1$). The only difference between these spin configurations is their statistic under the exchange of the two particles, which translates into the sign difference:
\begin{equation}
\ket{\varphi_S (K, -r)} = (-1)^S e^{i(K\cdot r)} \ket{\varphi_S (K, r)} \, .
\end{equation}
The action of the Hamiltonian Eq.~\ref{appeq:HamiltonianLargeU} on this basis is
\begin{align} 
& \mathcal{H}' \ket{\varphi_S (K, r)} = t \sum_{ \substack{j=1,2,3 \\ \epsilon=\pm}} \left[ 1 + e^{i \epsilon (K \cdot a_j)} \right] \ket{\varphi_S (K, r+\epsilon a_j)} \nonumber \\
& + \sum_{ \substack{j=1,2,3 \\ \epsilon=\pm}} \delta_{r,\epsilon a_j} \left[ V \ket{\varphi_S (K, \epsilon a_j)}  + \lambda \ket{\varphi_S (K, - \epsilon a_{j-\epsilon})}  \right] \nonumber \\ 
& + \lambda \sum_{ \substack{j=1,2,3 \\ \epsilon=\pm}} \delta_{r,\epsilon a_j} e^{i \epsilon (K\cdot a_{j-\epsilon})} \ket{\varphi_S (K, - \epsilon a_{j+\epsilon})} .
\end{align}
We then solve this equation numerically for large enough system sizes to extract the ground state energy in each spin sector. Our solution are shown in Fig.~\ref{fig_TwoBodyLattice} of the main text.

While our original model does not include any direct repulsion between $B$ sites on the honeycomb lattice, the two-particle bound state we have established is robust against longer range interactions. To study their effect, we further add non-local interaction between conduction electrons to the effective Hamiltonian $\mathcal{H}'$ and re-solve the two-particle problem.
We find that bound state is destroyed only when the the non-local repulsion becomes comparable to the exciton-induced short-range pairing interaction (which is much larger than the binding energy $\varepsilon_b$). If we take into account the direct Coulomb repulsion between nearest-neighbor $B$ sites $V'$, bound sate persists for $V' < 2\lambda - V$, or $0.25$eV when the parameters mentioned in the main text are used. 
If we include the long-range Coulomb interaction $\frac{e^2}{\epsilon r}$ fully, bound state exists for $\epsilon a > 86.4 \si{\angstrom}$ ($a$ is the lattice constant), which corresponds to $e^2/\epsilon a = 16.5$meV. Thus, in order for electron pairing to occur in the limit of vanishing doping, it is helpful to have a large $\epsilon$ which can result from dielectric screening by a different band.

\begin{widetext}

\section{Continuum Limit} \label{app:ContinuumLimit}

As shown in Fig.~\ref{fig_EffectiveModel}c, the kinetic part of the effective Hamiltonian Eq.~\ref{eq_effectivemodelBfermions} dominates over interactions. Thus, low-energy fermions live near the two degenerate minima of the single-particle dispersion relation located at the $K$ and $K'$ points in the Brillouin Zone. Our goal here is to derive an effective continuum field theory capturing the physics of the system when fermions remain close to these two valleys.

We start with the momentum representation of Eq.~\ref{eq_effectivemodelBfermions}
\begin{equation} \label{appeq_momentumspacehamiltonian}
\mathcal{H}_f = \sum_{k,\sigma} \varepsilon_k f_{k,\sigma}^\dagger f_{k,\sigma} + \frac{1}{2 N_s} \sum_{ \substack{k,q,p \\ \sigma,\sigma'} } V_{k,q} f_{k,\sigma}^\dagger f_{q+p,\sigma'}^\dagger f_{k+p,\sigma'} f_{q,\sigma}
\end{equation}
with $N_s$ the number of unit cells in the lattice, $\varepsilon_k = 2 t \sum_{j=1}^3 \cos(k\cdot a_j)$ and
\begin{equation} 
V_{k,q}  = U + 2 V \sum_j \cos[(k-q)\cdot a_j]  + 2 \lambda \sum_j ( e^{i k a_j + i q a_{j-1}} + e^{-i k a_{j-1} - i q a_j} ) + 2 \tilde{t} \sum_j [ \cos(k\cdot a_j)+\cos(q\cdot a_j) ]  .
\end{equation}
Due to the quadratic band dispersion near the $K$ and $K'$ points, low energy fermions acquire an effective mass $m = 2/(3 t_f a^2)$. They also carry an effective SU(4) $\{\uparrow K , \downarrow K , \uparrow K' , \downarrow K' \}$ that distinguishes both their spin and their valley degeneracy and enable contact interactions between fermions with the same spin, provided they have opposite valley index. 

Let us now focus on the scattering properties of these low energy fermions. Due to momentum conservation, two incoming low-energy fermions from the same valley can only scatter into a pair of fermions living in the same valley. The corresponding vertex interaction reads
\begin{equation}
V_c = V_{K,K} = V_{K', K'} = 6(V-\lambda-\tilde{t}) + U . 
\end{equation}
When the electrons are in opposite valley $K$ and $K'$, they can scatter to a pair in $K$ and $K'$ with the same valley preserving interaction strength $V_c$, or exchange valley to end up in $K'$ and $K$ through the vertex 
\begin{equation}
V_x = V_{K,K'} = V_{K',K} = 3 (4\lambda - 2\tilde{t} - V) + U .
\end{equation}

Introducing different fields for the two valleys 
\begin{equation}
f_{k,\sigma} = \begin{cases}
\psi_{k,\sigma,K} & \text{if } k \, \text{near } K \\
\psi_{k,\sigma,K'} & \text{if } k \, \text{near } K' 
\end{cases} ,
\end{equation}
and accounting for the $V_c$ and $V_x$ terms, we find that the following effective interacting Hamiltonian
\begin{equation} \begin{split}
\mathcal{H}_{\rm int} & = \frac{V_c}{N_s} \sum_{ \substack{k,q,p \\ V=K,K'} } \psi_{k,\uparrow,V}^\dagger \psi_{p-k,\downarrow,V}^\dagger \psi_{p-q,\downarrow,V} \psi_{q,\uparrow,V}  + \frac{V_c-V_x}{N_s} \sum_{ \substack{k,q,p \\ \sigma=\uparrow,\downarrow} } \psi_{k,\sigma,K}^\dagger \psi_{p-k,\sigma,K'}^\dagger \psi_{p-q,\sigma,K'} \psi_{q,\sigma,K} \\
& + \frac{V_c}{N_s} \sum_{ \substack{k,q,p \\ V=K,K'} } \psi_{k,\uparrow,V}^\dagger \psi_{p-k,\downarrow,\bar{V}}^\dagger \psi_{p-q,\downarrow,\bar{V}} \psi_{q,\uparrow,V} + \frac{V_x}{N_s} \sum_{ \substack{k,q,p \\ V=K,K'} } \psi_{k,\uparrow,V}^\dagger \psi_{p-k,\downarrow,\bar{V}}^\dagger \psi_{p-q,\downarrow,V} \psi_{q,\uparrow,\bar{V}} .
\end{split} \end{equation}
Let us rearrange these terms in terms of pair operators to make their physical meaning clearer. Valley-polarized spin-singlet electron pairs $S_{V} = f_{V, \downarrow} f_{V, \uparrow}$ with $V=K,K'$ only feel the valley conserving term and therefore exhibit repulsive interaction ($V_c>0$). When incoming electrons occupy opposite valleys, the ferromagnetic exchange leads to a total interaction strength $V_c + (-1)^S V_x$ depending on the total spin $S$ of the pair. As a consequence, the last spin-singlet valley-triplet channel $S_0 = ( f_{K',\downarrow} f_{K,\uparrow} - f_{K',\uparrow } f_{K,\downarrow} ) / \sqrt{2}$ is also repulsive ($V_c+V_x>0$). 
On the contrary, the three valley-singlet spin-triplet pair states, $T_{\sigma} = f_{K', \sigma} f_{K, \sigma}$ with $\sigma=\uparrow,\downarrow$ and $T_0 = (f_{K', \downarrow} f_{K, \uparrow} + f_{K', \uparrow} f_{K, \downarrow})/\sqrt{2}$, all display a low-energy interaction strength $V_c-V_x = 9(V - 2\lambda)$, which is negative for a wide range of parameter, as shown in Fig.~\ref{fig_RegionPairing}a.
To summarize, we can rewrite the different contact interaction terms as
\begin{equation} \label{appeq_fieldtheory}
\tilde{H} = \int {\rm d} x \sum_{\sigma, V} \psi_{\sigma, V}^\dagger \left[ \frac{-\nabla^2}{2m} \right] \psi_{\sigma, V}  + \int \frac{{\rm d}x}{\mathcal{A}} [ (V_c-V_x) (T_\downarrow^\dagger T_\downarrow + T_0^\dagger T_0 + T_\uparrow^\dagger T_\uparrow ) + V_c (S_{K'}^\dagger S_{K'} + S_K^\dagger S_K ) + (V_c+V_x) S_0^\dagger S_0 ] ,  
\end{equation} 
with $\mathcal{A} = 2 / \sqrt{3}a^2$ the Brillouin zone area.
This effective field theory describes a four-component Fermi liquid with repulsive interactions in the spin-singlet channel, owing to the large on-site interaction $U$ which appears in both $V_c$ and in $(V_c+V_x)$, and attractive interaction between fermions with total spin one when $V_c-V_x<0$ (see Fig.~\ref{fig_RegionPairing}a). This is the conclusion drawn in Eq.~\ref{eq_negative_coupling_constant} of the main text.

Alternatively, we can replace pair operators by more physical quantities, such as the total density on each valley $\rho_V = \psi_{\uparrow,V}^\dagger \psi_{\uparrow,V} + \psi_{\downarrow,V}^\dagger \psi_{\downarrow,V}$ and the total spin on each valley $\bm{s}_V = \psi_{\alpha,V}^\dagger \bm{\sigma}_{\alpha,\beta} \psi_{\beta,V}$. Together, they allow to represent the exchange term as
\begin{equation}
T_\downarrow^\dagger T_\downarrow + T_\uparrow^\dagger T_\uparrow + T_0^\dagger T_0 - S_0^\dagger S_0 = 2 \bm{s}_K \cdot \bm{s}_{K'} + \frac{1}{2} \rho_K \rho_{K'}  . 
\end{equation}
The valley conserving terms present in all interaction channels can be simply with the total density $\rho_{\rm tot} = \rho_K + \rho_{K'}$
\begin{equation}
T_\downarrow^\dagger T_\downarrow + T_\uparrow^\dagger T_\uparrow + T_0^\dagger T_0 + S_0^\dagger S_0 + S_K^\dagger S_K + S_{K'}^\dagger S_{K'} = \frac{1}{2} \rho_{\rm tot} (\rho_{\rm tot} - 1) . 
\end{equation}
Together, they allow to rewrite the interaction part of the continuum Hamiltonian as
\begin{equation}
\tilde{H}_{i} = \frac{1}{2 \mathcal{A}} \int {\rm d}x \left[ V_c \, n_{\rm tot} (n_{\rm tot} - 1) - V_x \left( 4 \bm{s}_K \cdot \bm{s}_{K'} + n_K n_{K'} \right) \right] .
\end{equation}
This forms makes clear the ferromagnetic interactions between opposite valleys, which are responsible for the formation of triplet pairs. Expanding the total density as a function of $\rho_K$ and $\rho_{K'}$, we find the three coupling constant given in Eq.~\ref{eq:CouplingConstantFieldTheory} of the main text
\begin{equation}
g_0 = V_c / (2\mathcal{A}), \quad g_1 = (2V_c - V_x)/(2\mathcal{A}) , \quad g_2 = -2V_x/\mathcal{A} .
\end{equation}

\end{widetext}

\section{Mean-field theory of superconductivity} \label{app:MeanFieldSC}

In this appendix, we carry out a mean-field treatment of the effective Hamiltonian Eq.~\ref{appeq_momentumspacehamiltonian} to investigate its superconducting behavior. With the mean-field substitution $f_{q',\sigma'} f_{q,\sigma} \simeq \delta_{q+q'} \langle f_{q',\sigma'} f_{q,\sigma} \rangle$, we get the following quadratic mean-field approximation:
\begin{align}
& \mathcal{H}_{\rm mf} = \sum_{k, \sigma} \xi_q f_{q,\sigma}^\dagger f_{q,\sigma} + \frac{1}{2} \sum_{k,\sigma,\sigma'} \left[ \tilde{\Delta}_{k, \sigma \sigma'} f_{k,\sigma}^\dagger f_{-k,\sigma'}^\dagger  + hc \right] , \nonumber \\  
& \tilde{\Delta}_{k,\sigma \sigma'} = - \frac{1}{N_s} \sum_q V_{k,q} \langle f_{q,\sigma} f_{-q,\sigma'} \rangle ,
\end{align}
with $\xi_k = \xi_{-k} = \varepsilon_k-\mu$ and $\mu$ the chemical potential. 
It can be rewritten as a sum over a halved Brillouin Zone (denoted with primed sums and products below):
\begin{equation}
\mathcal{H}_{\rm mf}  = \sum_{k}^{'}  \begin{bmatrix} f_k^\dagger & f_{-k} \end{bmatrix} \begin{bmatrix} \xi_k & \Delta_k \\ \Delta_k^\dagger & - \xi_k \end{bmatrix} \begin{bmatrix} f_k \\ f_{-k}^\dagger \end{bmatrix} \, .
\end{equation}
The order parameters have been gathered in a $2\times2$ matrix 
\begin{equation}
\Delta_k = \frac{\tilde{\Delta}_k - \tilde{\Delta}_{-k}^T}{2} = \frac{-1}{N_s} \sum_q \Re(V_{k,q}) \langle f_{q,\sigma} f_{-q,\sigma'} \rangle ,
\end{equation}
and should be computed self-consistently.

\subsection{Pairing symmetries}

The explicit expression of $V_{k,q}$ allows to decompose this order parameter into spin-singlet and spin-triplet components
\begin{equation} \label{eq:MeanField_OrderParamSymmetry}
\Delta_k = \Delta_s' + \sum_{j=1}^3 \Delta_j^s \cos(k\cdot a_j) + \Delta_j^t \sin(k\cdot a_j) ,
\end{equation}
which respectively read:
\begin{align}
\Delta_s' & = \frac{-1}{N_s} \sum_q \left[ U + 2\tilde{t} (c_1+c_2+c_3) \right] \langle f_{q,\sigma} f_{-q,\sigma'} \rangle , \notag \\
\Delta_j^s & = \frac{-2}{N_s} \sum_q \left[ \lambda(c_{j-1}+c_{j+1}) + V c_j + \tilde{t} \right] \langle f_{q,\sigma} f_{-q,\sigma'} \rangle , \notag \\
\Delta_j^t & = \frac{ 2}{N_s} \sum_q \left[ \lambda(s_{j-1}+s_{j+1}) - V s_j \right] \langle f_{q,\sigma} f_{-q,\sigma'} \rangle ,
\end{align}
with $c_j = \cos(q\cdot a_j)$ and $s_j = \sin (q\cdot a_j)$. 
We can further split these order parameters in terms of irreducible representation of $C_{3v}$ that they represent on the triangular lattice. 
For the singlet and triplet case, there are two one-dimensional irrep, only one of which can be obtained because of the particular form of $V_{k,q}$ and one two dimensional irrep. The two former measure the strength of s-wave and f-wave pairing 
\begin{equation}
\Delta_s' , \quad 
\Delta_{s/f} = \frac{1}{3} \left[ \Delta_1^{s/t}+\Delta_2^{s/t}+\Delta_3^{s/t} \right] ,
\end{equation}
while the two dimensional irrep are related to d-wave and p-wave pairing 
\begin{align}
\Delta_{d_{x^2-y^2}/p_x} & = \frac{1}{6} \left[ \Delta_1^{s/t} + \Delta_2^{s/t} - 2\Delta_3^{s/t} \right]  , \\
\Delta_{d_{xy}/p_y} & = \frac{1}{2} \left[ \Delta_1^{s/t}-\Delta_2^{s/t} \right]  .
\end{align}
The inverse transformations are 
\begin{align}
\Delta_1 & = \Delta_{s/f} + \Delta_{d_{x^2-y^2}/p_x} + \Delta_{d_{xy}/p_y} , \notag\\
\Delta_2 & = \Delta_{s/f} + \Delta_{d_{x^2-y^2}/p_x} - \Delta_{d_{xy}/p_y} , \notag\\
\Delta_3 & = \Delta_{s/f} - 2 \Delta_{d_{x^2-y^2}/p_x} .
\end{align}
Finally, singlet pairs cannot be of equal spin, and we can therefore express them as a scalar times the $2\times2$ matrix $i\sigma_y$, \textit{e.g.} $\Delta_s = d_0 (i\sigma_y)$ with $\delta_s$ a complex number.
Triplet on the other hand, take the form of a Pauli vector multiplied by $i\sigma_y$, \textit{e.g.} $\Delta_f = (\bm{d}_f \cdot \bm{\sigma})  (i\sigma_y)$.

Our mean-field treatment relies on the self-consistent computation of four scalars related to singlet pairing in s-wave ($\Delta_s$, $\Delta_s'$) of d-wave ($\Delta_{d_{x^2-y^2}}$, $\Delta_{d_{xy}}$), and three vectors describing f-wave ($\Delta_f$) or p-wave ($\Delta_{p_x}$, $\Delta_{p_y}$) pairs. The corresponding self-consistent equations become
\begin{align} \label{appeq_selfconsistentgapeq}
&\Delta_s' = \frac{-1}{N_s} \sum_q \left[ U + 2\tilde{t} (c_1+c_2+c_3) \right] \langle f_{q,\sigma} f_{-q,\sigma'} \rangle , \notag \\ 
& \Delta_s  = \frac{-2}{N_s} \sum_q \left[ \tilde{t} + \frac{2\lambda + V}{3} (c_1+c_2+c_3) \right] \langle f_{q,\sigma} f_{-q,\sigma'} \rangle , \notag \\
&\Delta_{p_x} = \frac{-2 (\lambda + V)}{N_s} \sum_q \frac{s_1+s_2-2s_3}{6} \langle f_{q,\sigma} f_{-q,\sigma'} \rangle , \notag \\
&\Delta_{p_y} = \frac{-2 (\lambda + V)}{N_s} \sum_q \frac{s_1-s_2}{2} \langle f_{q,\sigma} f_{-q,\sigma'} \rangle , \notag \\
&\Delta_{d_{x^2-y^2}} = \frac{2 (\lambda - V)}{N_s} \sum_q \frac{c_1+c_2-2c_3}{6} \langle f_{q,\sigma} f_{-q,\sigma'} \rangle , \notag \\ 
& \Delta_{d_{xy}} = \frac{2 (\lambda - V)}{N_s} \sum_q \frac{c_1-c_2}{2} \langle f_{q,\sigma} f_{-q,\sigma'} \rangle , \notag \\
&\Delta_f  = \frac{2 (2\lambda - V)}{N_s} \sum_q \frac{s_1+s_2+s_3}{3} \langle f_{q,\sigma} f_{-q,\sigma'} \rangle .
\end{align}

\subsection{Self-consistent conditions}

The mean field quadratic Hamiltonian can be diagonalized by a Bogoliubov transformation. 
Writing the hermitian matrix 
\begin{equation}
\Delta_q \Delta_q^\dagger = a_0 + \bm{a}\cdot \bm{\sigma}
\end{equation}
as a Pauli vector, the eigen-energies read
\begin{equation}
E_{q,\pm} = \sqrt{\xi_q^2 + a_0 \pm |\bm{a} |} .
\end{equation}
The corresponding eigenvectors lead to the following expression for the anomalous correlators
\begin{equation}  \label{appeq_meanfield_selfconscorrel}
\langle f_q f_{-q}^T \rangle = \left[ g_q^+ + g_q^- \frac{\bm{a}\cdot\bm{\sigma}}{|a|} \right] \frac{\Delta_q}{2} , 
\end{equation}
with
\begin{equation} 
g_q^\pm = \frac{g(E_{q,+}) \pm g(E_{q,-})}{2} , \; g(E) = \frac{\tanh(\beta E/2)}{E} .
\end{equation}
When $\Delta_q \Delta_q^\dagger$ is simply proportional to the identity, for instance when pairing occurs for spin-singlet, $E_{q,+} = E_{q,-} = E_q$ and the previous expression simply becomes $\langle f_q f_{-q}^T \rangle = g(E_q) \Delta_q/2$.

\subsection{Critical temperature}

Solving the self-consistent relations of Eq.~\ref{appeq_selfconsistentgapeq} with the help of Eq.~\ref{appeq_meanfield_selfconscorrel} allows to determine the nature of the superconducting state. 
We now consider each pairing channel separately to check a superconducting phase can fully form. 

\subsubsection{s-wave}

The possibility of an s-wave SC order can be ruled out because $U$ is much larger than all the other terms scaling as $t_0^2/\Delta$. The coupled gap equations for $\Delta_s$ and $\Delta_s'$ linearized near $T_c$ read
\begin{equation} 
\begin{bmatrix} \Delta_s' \\ \Delta_s \end{bmatrix} = \begin{bmatrix} U I_0 + 2\tilde{t} I_1 & U I_1 + 2\tilde{t} I_2 \\
2\tilde{t} I_0 + \frac{2(2\lambda+V)}{3} I_1 & 2\tilde{t} I_1 + \frac{2(2\lambda+V)}{3} I_2 \end{bmatrix}  \begin{bmatrix} \Delta_s' \\ \Delta_s \end{bmatrix}  ,
\end{equation}
where $I_k = - \sum_q \tanh(\beta E_q/2) (c_1+c_2+c_3)^k / (2 N_s E_q)$ and $\beta$ the inverse temperature. The matrix in the previous equation must have at least one eigenvalue equal to one for the system to exhibit s-wave symmetry. However, to leading order in $U$, this requires to have 
\begin{equation}
U \left[ \frac{2(2\lambda+V)}{3} (I_0 I_2 - I_1^2) - I_0 \right] = 0 . 
\end{equation}
This equation has no solution because $(-I_0)>0$ has the same sign as $I_0 I_2 - I_1^2 > 0$ (due to Cauchy-Schwarz iequality).
Thus, singlet pairing does not occur happen. It could nevertheless appear for smaller ratios $\Delta/t_0$ where our perturbation theory breaks down.

\subsubsection{p-wave}

The possibility of p-wave pairing can be ruled out as well. Indeed, let's assume a p-wave SC order and compute the critical temperature $T_c$ of that state. 
Linearizing the gap equation, such that $\bm{a} \simeq \bm{0}$ and $E_{q,\pm} \simeq |\xi_q|$, we find the coupled equations
\begin{widetext}
\begin{equation}
\begin{bmatrix} \Delta_{p_x} \\  \Delta_{p_y} \end{bmatrix} = - \frac{V + \lambda}{N_s} \sum_q \frac{\tanh(\beta E_q /2)}{E_q} \begin{bmatrix} (s_1+s_2-2s_3)^2/6 & (s_1+s_2-2s_3)(s_1-s_2)/6 \\ (s_1-s_2)(s_1+s_2-2s_3)/2 &  (s_1 - s_2)^2/2 \end{bmatrix} \begin{bmatrix} \Delta_{p_x} \\ \Delta_{p_y} \end{bmatrix} \, .
\end{equation}
\end{widetext}
Noting that $E_q = |\xi_q|$ is $C_3$ invariant, while the off-diagonal terms of the equation are not, we can rewrite the diagonal terms as:
\begin{equation}
\frac{-6}{V + \lambda} = \frac{1}{N_s} \sum_q \frac{\tanh(\beta E_q /2)}{E_q} \sum_j (s_j-s_{j+1})^2 ,
\end{equation}
which does not have any solution since the left and right hand sides have opposite signs.

\subsubsection{d- and f-wave}

\begin{figure*}
\centering
\includegraphics[width=\textwidth]{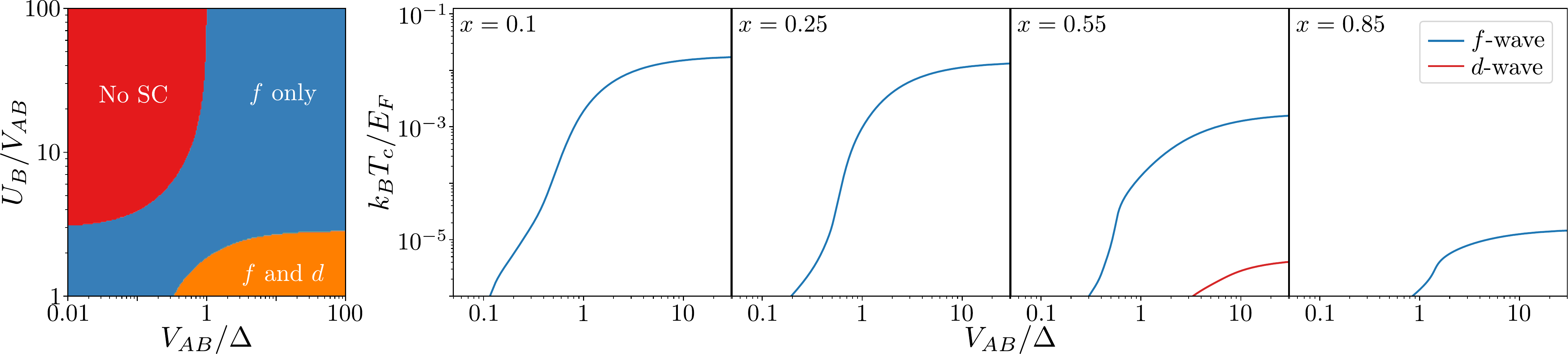}
\caption{ 
f-wave pairing only occurs if $V>2\lambda$ (blue), and d-wave pairing may compete with the latter if $V>\lambda$ (orange). The numerically extracted $T_c$ for $U_B=2V_0$ show that f-wave is either the only possible pairing channel or strongly dominate over d-wave symmetry. 
}
\label{appfig_Tcinallpairingchannel}
\end{figure*}

We can similarly derive an implicit equation for the critical temperature in the d-wave channel
\begin{equation} \label{eq:MeanField_DwavePairingTC}
\frac{6}{\lambda - V} = \frac{1}{N_s} \sum_q \frac{\tanh(\beta E_q /2)}{E_q} \sum_j (c_j-c_{j+1})^2 . 
\end{equation}
Contrary to p-wave pairing, $\lambda - V$ can be positive for $U_B$ not too large and $V_0/\Delta$ large enough. We highlight the region where $\lambda > V$ in Fig.~\ref{appfig_Tcinallpairingchannel}.

The same situation arise for f-wave pairing, whose critical temperature follows from
\begin{equation} \label{eq:MeanField_FwavePairingTC}
\frac{3}{2\lambda - V} = \frac{1}{N_s} \sum_q (s_1+s_2+s_3)^2 \frac{\tanh(\beta E_q/2)}{E_q} .
\end{equation}
As explained in the main text, $2 \lambda > V$ for most choice of parameter, leading to a superconducting order with f-wave symmetry in most cases.

When f- and d- wave superconductivity may coexist, we generically expect the f-wave critical temperature to dominate due to a larger interaction strength $2 \lambda - V > \lambda-V$. This is confirmed by our numerical solution of Eqs.~\ref{eq:MeanField_FwavePairingTC} and~\ref{eq:MeanField_DwavePairingTC}, where we observe that the f-wave critical temperature dominates. At large doping concentration $x>0.5$, the f-wave superconducting order parameter exhibits nodes along the $\Gamma-M$ lines, while d-wave paired state can avoid the presence of nodes. The latter may therefore be energetically favored. To investigate this competition, we solve the zero temperature mean-field equations assuming $\Delta_f$, $\Delta_{d_{x^2-y^2}}$ and $\Delta_{d_{xy}}$ nonzero. Our results presented in Fig.~\ref{appfig_Tcinallpairingchannel} show that the f-wave superconducting state still wins over d-wave order, despite its nodes, even when $x > 0.5$. Numerical data right at the van Hove singularity $x = 0.5$ show a more subtle competition between the two orders. At that point, other instabilities, \textit{e.g.} charge density wave, may arise and our mean-field treatment should be complemented with more precise analytical tools. We leave these details for a future work, simply noting for now that f-wave superconductivity largely dominate in the entire phase diagram of our model, except maybe at the van Hove singularity.

\subsection{Zero temperature superconducting gap}

As discussed in the main text, the superconducting gap for f-wave and spin-triplet symmetry may be written as $\Delta_k = (\bm{d}_k \cdot \bm{\sigma}) (i\sigma_y)$ with $\bm{d}_k = [ \sum_j \sin(k\cdot a_j) ] \bm{d}_f$ and $\bm{d}_f$ a constant vector that only depends on temperature. Owing to the spin-rotation symmetry of our model, we can choose $\bm{d}_f$ to lie in the $xy$ plane at on point of the Brillouin zone. Moreover, we can eliminate any phase difference between $\Delta_k^{\uparrow\uparrow}$ and $\Delta_k^{\downarrow\downarrow}$, which yields $\bm{d}_f = e^{i\theta} (d_x, id_y, 0)^T$ with $d_{x,y}$ real. With this order parameter, eigen-energies of the mean-field Hamiltonian take the rather simple form
\begin{equation}
E_{k,\pm} = \sqrt{\xi_k^2+ s_k^2 |d_x \pm d_y|^2} .
\end{equation}
The self-consistent equation for such symmetry (see Eq.~\ref{appeq_selfconsistentgapeq}) gives 
\begin{equation} \label{eq:selfconsistentdxdy}
\begin{bmatrix} d_x \\ d_y \end{bmatrix} = \frac{2\lambda - V}{3} \begin{bmatrix} a^+ & a^- \\ a^- & a^+ \end{bmatrix} \begin{bmatrix} d_x \\ d_y \end{bmatrix} ,
\end{equation}
with 
\begin{equation}
a^\pm = \frac{1}{N_s} \sum_q s_q^2 g_q^\pm , \quad s_q = \sum_j \sin (q\cdot a_j) .
\end{equation}

\begin{figure}
\centering
\includegraphics[width=\columnwidth]{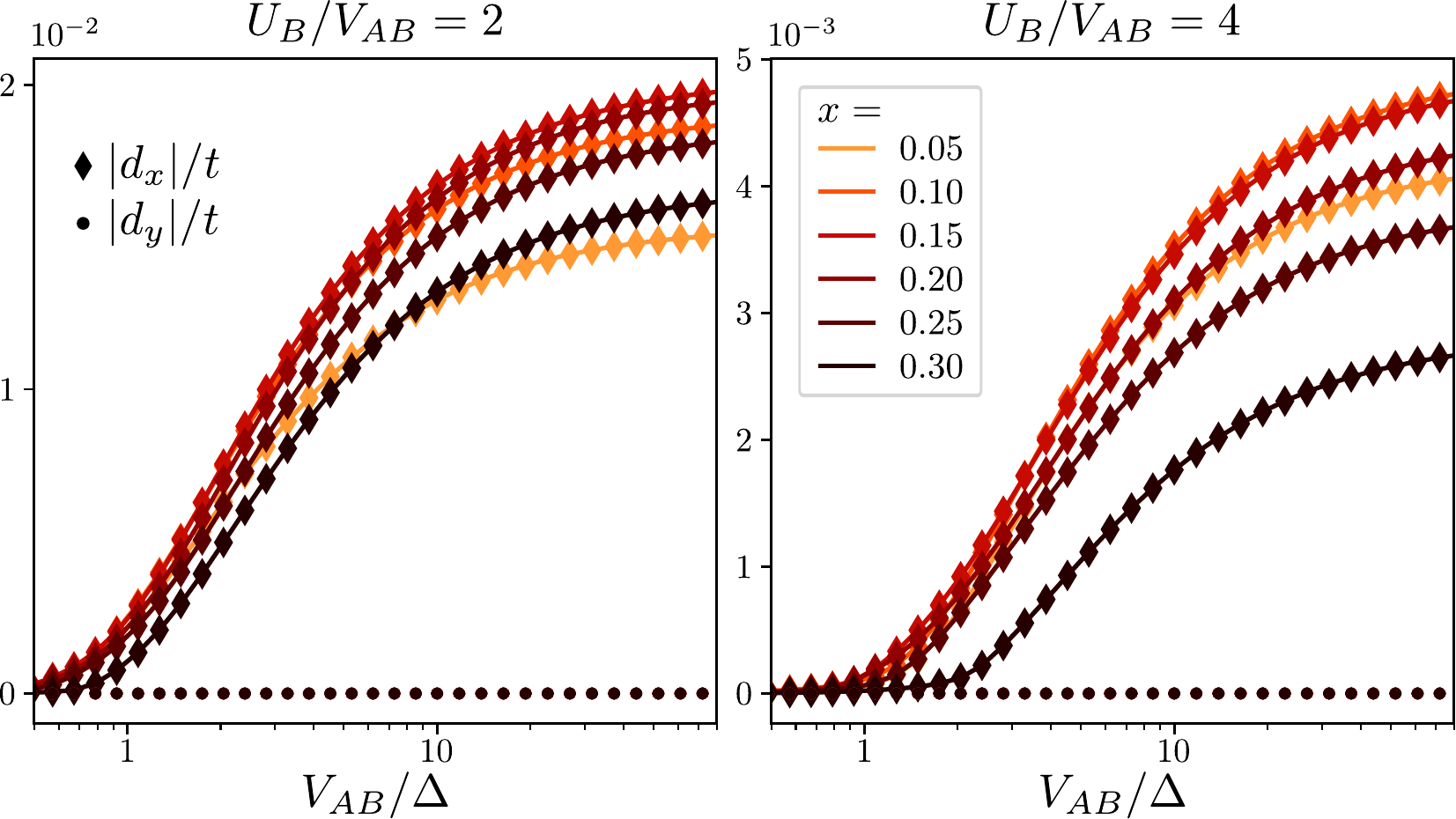}
\caption{ 
Zero temperature order parameters $d_x$ and $d_y$ as a function of $V_0/\Delta$ for different doping concentration $x$ and ratios of $U_B / V_0$. In all cases considered, the superconducting state is polar, \textit{i.e.} $d_y=0$. 
}
\label{appfig_polarstate}
\end{figure}

To have a non-zero solution, we must have
\begin{equation}
\left| a^+ - \frac{3}{2\lambda - V} \right| = |a^-| .
\end{equation}
The solution of this equation with the large total gap $|\bm{d}|$ necessarily has $a^-=0$. Because $s_q^2 g_q^-$ has the same sign in the entire Brillouin zone, this implies that $E_{q,+} = E_{q,-}$ for all momenta. Hence, $d_x=0$ or $d_y=0$ in the entire Brillouin zone. In both cases, the net spin carried by the Cooper pairs is zero~\cite{leggett1975theoretical}:
\begin{equation}
\langle \bm{S} \rangle = \langle i \bm{d} \times \bm{d}^* \rangle = 0 .
\end{equation}
As a consistency check, we have numerically solved the self-consistent Eq.~\ref{eq:selfconsistentdxdy} and reached the same conclusion. For all parameters considered, our results show $d_y=0$, as exemplified in Fig.~\ref{appfig_polarstate} for $U_B/V_0=2$ and $U_B/V_0=4$.

\section{Weakly interacting regime} \label{app:WeakInteraction}
\subsection{Explicit unitary transformation}

We start by isolating the band mixing (or off-diagonal) interaction elements from the others $\mathcal{V} = \mathcal{V}_{\rm od} + \mathcal{V}_{\rm d}$. 
More explicitly, we write $\mathcal{V}_{\rm od} = \frac{1}{N_s} \sum_{C(1234)} V_{43}^{21} \delta_{43}^{21}  c_4^\dagger c_3^\dagger c_2 c_1$, with $C(1234)$ restricting the sum to terms satisfying either $b_1b_2b_3b_4 = -1$, or having $b_1=b_2$ and $b_3=b_4$ with $b_1\neq b_2$.

To eliminate these band mixing interaction terms, we use a Schrieffer-Wolff transformation $\mathcal{H}' = e^S \mathcal{H} e^{-S}$, with $S$ anti-hermitian. 
This unitary transformation can be carried order by order in the small parameter $|\mathcal{V}|/\Delta$, and we write $S = S_1+S_2+\cdots$ with $S_n = \mathcal{O}(|\mathcal{V}|^n / \Delta^n)$. 
Requiring 
\begin{equation}
[ H_0 , S_1 ] = \mathcal{V}_{\rm od} ,
\end{equation}
gets rid of the direct band-mixing terms in $\mathcal{H}'$:
\begin{align} 
\mathcal{H}' & = \mathcal{H} + [S, \mathcal{H}] + \frac{1}{2} [S, [S, \mathcal{H}]] + \mathcal{O}(S^2 \mathcal{H}) \\
& = \mathcal{H}_0 + \mathcal{V}_{\rm d} + \left[ S_1 , \mathcal{V}_{\rm d} +\frac{\mathcal{V}_{\rm od}}{2} \right] + [S_2, H_0] +  \mathcal{O}\left( \frac{|\mathcal{V}|^3}{\Delta^2} \right) .  \notag
\end{align}
A possible $S_1$ satisfying this condition is
\begin{equation} \label{appeq:SchriefferWolffExplicit}
S_1 = \sum_{C(1234)} \frac{V_{43}^{21} \delta_{43}^{21} }{\varepsilon_4 + \varepsilon_3 - \varepsilon_2 - \varepsilon_1 }  c_4^\dagger c_3^\dagger c_2 c_1  .
\end{equation}
We can then choose  $[S_2, H_0]$ to remove the band mixing terms of $[S_1, \mathcal{V}_{\rm d}+\mathcal{V}_{\rm od}/2]$, and so on. 
Note that, because $\mathcal{V}_{\rm d}$ contains no band mixing terms, such that $[S_1, \mathcal{V}_{\rm d}]$ is purely off diagonal and therefore completely eliminated by $[S_2, H_0]$. 
The second order corrections read $[S_1, \mathcal{V}_{\rm od}] / 2$, which as promised are of order $|\mathcal{V}|^2 / \Delta$ due to the special form of $S_1$. 

Finally, the effective Hamiltonian for doped charge is obtained by projecting $\mathcal{H}'$ to the subspace where the lower band is fully filled, which amounts to pairing up lower band indices as $c_a^\dagger c_b \to \tilde{\delta}_{a,b}$.

\subsection{Leading corrections}

To make analytical progress, we need the explicit form of the dispersion and scattering vertex. 
The single particle energy dispersion reads
\begin{equation}
\varepsilon_{\vec{k},\pm,\sigma} = \varepsilon_{\vec{k},\pm} = \pm  \sqrt{(\Delta_0/2)^2 + |t_0 f(\vec{k})|^2} , 
\end{equation}
with $f(\vec{k}) = \sum_{j=1}^3 e^{i(\vec{k} \cdot \vec{u}_j)}$, and $\vec{u}_{j=1,2,3}$ the vectors connecting $B$ sites to their three nearest neighbors. 
The corresponding Bloch eigenvectors are
\begin{equation}
\Psi_{\vec{k},\pm} = \frac{1}{\sqrt{2\varepsilon_{\vec{k},+} (\varepsilon_{\vec{k},+} \pm \Delta_0/2)}} \begin{bmatrix} \mp t_0 f(\vec{k}) \\ \varepsilon_{\vec{k},+} \pm \Delta_0/2
\end{bmatrix} ,
\end{equation}
They allow us to write the interaction vertex in the band basis as
\begin{equation} \label{appeq:DefinitionVertex} \begin{split}
& V_{43}^{21} =  V_0 f(\vec{k}_4-\vec{k}_1) \left[ \Psi_{\vec{k}_4,b_4}^{A} \Psi_{\vec{k}_3,b_3}^{B} \right]^*  \Psi_{\vec{k}_2,b_2}^{B} \Psi_{\vec{k}_1,b_1}^{A}  \\
& + \!\!\! \sum_{\tau=A/B} \!\!\! \frac{U_\tau}{2} \delta_{(\sigma_4=\sigma_1) \neq (\sigma_3=\sigma_2)} \left[ \Psi_{\vec{k}_4,b_4}^{\tau} \Psi_{\vec{k}_3,b_3}^{\tau} \right]^*  \Psi_{\vec{k}_2,b_2}^{\tau} \Psi_{\vec{k}_1,b_1}^{\tau} .
\end{split} \end{equation}

\begin{figure}
\centering
\includegraphics[width=0.85\columnwidth]{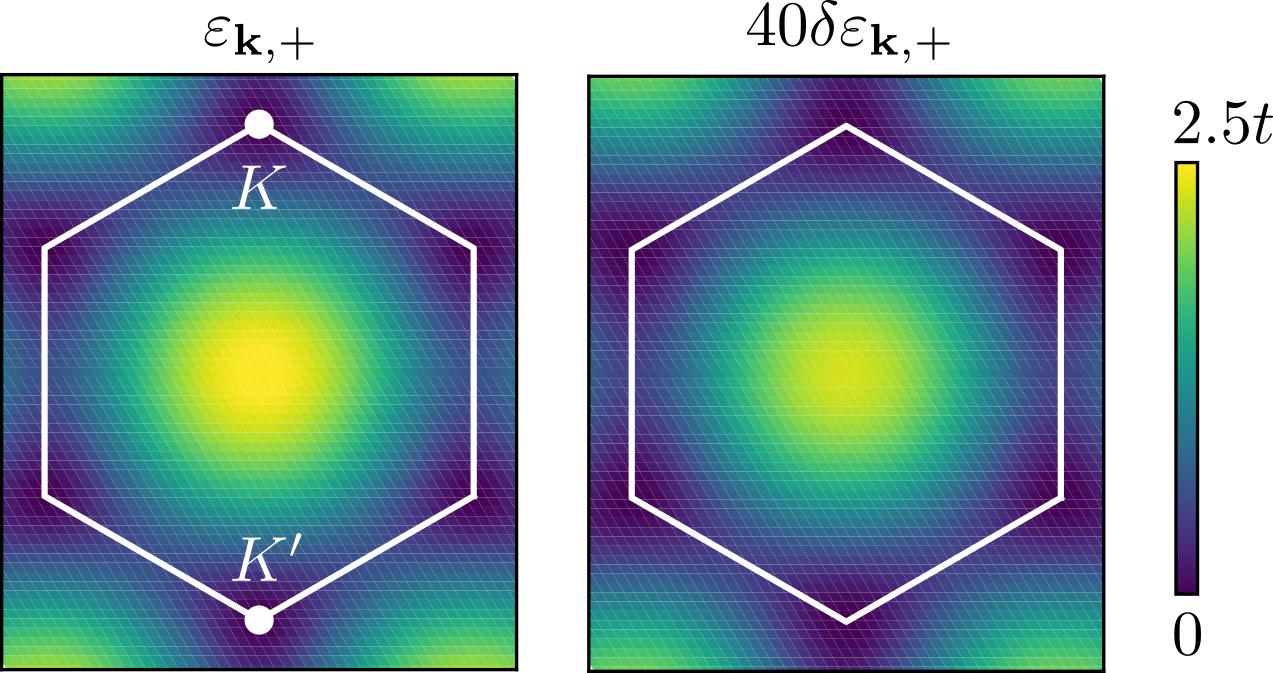}
\caption{ 
Bare dispersion $\varepsilon_{\vec{k},+}$ and its leading correction $\delta \varepsilon_{\vec{k},+}$ (multiplied 40 times for visibility) in the weakly interacting limit, shown for $\Delta_0 = t = 2U_A = 2U_B = 10V_0$. 
}
\label{fig_BandDispersion}
\end{figure}

The leading corrections to the band dispersion come from $\mathcal{V}_{\rm d}$ and are given by Eq.~\ref{eq:WeakIntEffMass} in the main text. 
The Hartree-like part, corresponding to the first line of Eq.~\ref{eq:WeakIntEffMass}, takes the explicit form
\begin{equation} \begin{split}
\delta \varepsilon_{\vec{k},+}^{\rm H} & = (6V_0 C_- + U_A C_+) \frac{\varepsilon_{\vec{k},+} - \Delta_0/2}{2\varepsilon_{\vec{k},+} } \\ & + (6V_0 C_+ + U_B C_-) \frac{\varepsilon_{\vec{k},+} + \Delta_0/2}{2\varepsilon_{\vec{k},+} } , 
\end{split} \end{equation}
with $C_\pm = (2 N_s)^{-1} \sum_{\vec{q}} (\varepsilon_{\vec{q},+} \pm \Delta_0/2) /\varepsilon_{\vec{q},+}$. 
The Fock-like term, second line of Eq.~\ref{eq:WeakIntEffMass}, reads
\begin{equation}
\delta \varepsilon_{\vec{k},+}^{\rm F} = \frac{t_0^2 V_0}{2 N_s} \Re\left[ \frac{f^*(\vec{k})}{\varepsilon_{\vec{k},+}} \sum_{\vec{q}} f(\vec{k}-\vec{q}) \frac{f(\vec{q})}{\varepsilon_{\vec{q},+}} \right] . 
\end{equation}
These corrections are plotted together with the bare band dispersion for $\Delta_0 = t = 2U_A = 2U_B = 10V_0$ in Fig.~\ref{fig_BandDispersion}, where we observe that they both admit degenerate minima at the $K$ and $K'$ points. 

When $t_0 \ll \Delta_0$, we can approximate $\varepsilon_{\vec{k},+} \simeq \Delta_0/2 + |t_0 f(\vec{k})|^2/\Delta_0$ and expand these corrections in powers of $t_0/V_0$. 
Without too much difficulty, we end up with 
\begin{align}
\delta \varepsilon_{\vec{k},+}^{\rm H} & = \frac{|t_0 f(\vec{k})|^2}{\Delta_0^2} (U_A - 6V_0) , \\
\delta \varepsilon_{\vec{k},+}^{\rm F} & = \frac{2 t_0^2 V_0}{\Delta_0^2 N_s} \Re\left[ f^*(\vec{k}) \sum_{\vec{q}} f(\vec{k}-\vec{q}) f(\vec{q}) \right] , \notag
\end{align}
up to an overall global constant. 
These expressions allow to obtain the effective mass by expanding around the $K$ or $K'$ point. 
For that purpose, we recall $f(K+\vec{k}) \simeq \sqrt{3} (k_x+ik_y) a / 2$ and furthermore find that 
\begin{align} 
& \sum_{\vec{q}} f(K+\vec{k}-\vec{q}) f(\vec{q}) = \sum_{\vec{q}} f(\vec{k}-\vec{q}) f(\vec{q}+K) \notag \\ & \simeq \sum_{\vec{q}} f(-\vec{q}) f(\vec{q}+K) + \vec{k} \cdot \sum_{\vec{q}} (\nabla f)(-\vec{q}) f(\vec{q}+K ) \notag \\ & = 0  + \sqrt{3} (k_x+ik_y) a / 2,
\end{align}
where the last equality is easy to check numerically. 
We end up with 
\begin{equation}
(\varepsilon + \delta\varepsilon^{\rm H} + \delta\varepsilon^{\rm F})_{K + \vec{k},+} \simeq \frac{3 t_0^2 a^2 |\vec{k}|^2}{4 \Delta_0^2} (\Delta_0 + U_A  - 4V_0) ,
\end{equation}
yielding the effective mass given in Eq.~\ref{eq:WeakIntEffMass}.

We now turn to the corrections to the two-body scattering vertex, which are contained in the second order term $X = [S_1, \mathcal{V}_{\rm od}]/2$. 
Direct evaluation of the commutator using the explicit expression of $S_1$ (Eq.~\ref{appeq:SchriefferWolffExplicit}) gives 
\begin{equation}
X = \frac{1}{N_s} \sum_{123456} \Gamma_{654}^{321} \delta_{654}^{321} c_6^\dagger c_5^\dagger ( 2 c_4^\dagger c_3 - \delta_{3,4} ) c_2 c_1 , 
\end{equation}
where the indices $653$ and $421$ originates from the same interaction elements and thus satisfy the conditions given above for the elements of $\mathcal{V}_{\rm od}$ and $S_1$. 
The three-body tensor $\Gamma$ takes the form
\begin{align} \label{appeq:ExpressionOfGammaWeakInt}
\Gamma_{654}^{321} = \frac{1}{N_s} \sum_0 & \delta_{65}^{30} \delta_{40}^{21} (V_{65}^{03} - V_{65}^{30}) (V_{04}^{21} - V_{40}^{21})  \\
& \!\!\!\!\! \times \left[ \frac{1}{\varepsilon_6+\varepsilon_5 - \varepsilon_3-\varepsilon_0} + \frac{1}{\varepsilon_2+\varepsilon_1 - \varepsilon_4 - \varepsilon_0} \right] . \notag
\end{align}
For later use, we also define $\tilde{\Gamma}$ and $\hat{\Gamma}$, which have the same explicit representation except that the sum over $0$ is restricted to states in lower band $b_0 = -$ and the upper band $b_0=+$, respectively. 
To find the two-body corrections from $X$, we can select the terms where two of the indices belong to the lower band and contract them. 
Considering all possible pairs compatible with the constraints on $653$ and $421$, we end up with 
\begin{align} 
\delta V_{43}^{21} & = 2 \sum_{i , b_i = -} \Gamma_{43i}^{i21} + \hat{\Gamma}_{i43}^{2i1} + \hat{\Gamma}_{i43}^{21i} + \hat{\Gamma}_{4i3}^{2i1} + \hat{\Gamma}_{4i3}^{21i} \\
& + 2 \sum_{i , b_i = -} \tilde{\Gamma}_{i43}^{i21} + \tilde{\Gamma}_{4i3}^{i21} + \tilde{\Gamma}_{43i}^{2i1} + \tilde{\Gamma}_{43i}^{21i} - \sum_i \tilde{\Gamma}_{43i}^{i21} .  \notag
\end{align}

This expression greatly simplifies when we assume that the four momenta $\vec{k}_{1,2,3,4}$ are equal to $K$ or $K'$, as we do to determine the effective interaction strength in the spin-triplet valley singlet channel $U_0$ (see main text). 
Because $\Psi_{K/K',+}^A = 0$, any term of the form $\Gamma_{654}^{321}$ with $6=(K/K',+)$ or $1=(K/K',+)$ appearing in $U_0$ vanishes (see Eq.~\ref{appeq:ExpressionOfGammaWeakInt}). 
This completely removes any contribution from on-site interactions on $A$ sites $U_A$. 
Let us first focus on the $V_0^2$ contributions and set $U_B=0$ for a moment. 
This largely simplifies the expression of $U_0$, which now reads 
\begin{equation} \begin{split}
U_0^{(1)} = 2 \sum_{q} & \hat{\Gamma}_{(q-)(K+)(K'+)}^{(K'+)(K+)(q-)}  + \hat{\Gamma}_{(q-)(K'+)(K+)}^{(K+)(K'+)(q-)} \\ 
& - \hat{\Gamma}_{(q-)(K+)(K'+)}^{(K+)(K'+)(q-)} - \hat{\Gamma}_{(q-)(K'+)(K+)}^{(K'+)(K+)(q-)} . 
\end{split} \end{equation}
The two first terms of the sum involve interaction coefficients with momentum transfer $K$ and $K'$, respectively, which makes them vanish as $f(K)=f(K')=0$ (see Eq.~\ref{appeq:DefinitionVertex}). 
The other terms contribute equally, and we finally obtain
\begin{equation}
U_0^{(1)} = - \frac{36 t_0^2 V_0^2}{N_s} \sum_{\vec{q}} \frac{|f(\vec{q})|^2}{(2\varepsilon_{\vec{q},+})^3} . 
\end{equation}
It is not difficult to check that the contribution proportional to $U_B^2$ vanishes, and we now turn to the crossed $V_0 U_B$ corrections. 
The calculation proceeds in a similar way, except that one of the interaction element $3V_0$ is replaced by $U_B$ and the signs needs to be flipped, such that the exchange part gives 
\begin{equation}
U_0^{(2)} = \frac{12 t_0^2 V_0 U_B}{N_s} \sum_{\vec{q}} \frac{|f(\vec{q})|^2}{(2\varepsilon_{\vec{q},+})^3} . 
\end{equation}
For $t_0 \ll \Delta_0$, we use $\sum_{\vec{q}} |f(\vec{q})|^2 = 3N_s$ to find the simpler form 
\begin{equation}
U_0 = \frac{36 t_0^2 V_0 (U_B - 3V_0)}{\Delta_0^3} , 
\end{equation}
which -- remarkably -- exactly match the result of the kinetic expansion in the weakly interacting regime
\begin{equation}
U_0^{\rm KE} = V_c - V_x \stackrel{(U_A, U_B, V_0 \ll \Delta_0)}{\simeq} \frac{36 t_0^2 V_0 (U_B - 3V_0)}{\Delta_0^3} . 
\end{equation}

\end{document}